\documentclass{article}

\usepackage[preprint]{neurips_2026}

\usepackage[utf8]{inputenc} % allow utf-8 input
\usepackage[T1]{fontenc}    % use 8-bit T1 fonts
\usepackage{hyperref}       % hyperlinks
\usepackage{url}            % simple URL typesetting
\usepackage{booktabs}       % professional-quality tables
\usepackage{amsfonts}       % blackboard math symbols
\IfFileExists{nicefrac.sty}{\usepackage{nicefrac}}{}
\IfFileExists{microtype.sty}{\usepackage{microtype}}{}
\usepackage{xcolor}         % colors
\usepackage{enumitem}       % itemize / enumerate options

\hypersetup{hidelinks, pdfauthor={Mateo Pechon-Elkins, Jon Chun}, pdftitle={Auditing Game-Theoretic Measures of Strategic Reasoning in LLMs}, pdfsubject={}, pdfkeywords={}, pdfproducer={}, pdfcreator={LaTeX with hyperref}}

\usepackage[american]{babel}

\usepackage{mathtools} % extends amsmath
\usepackage{amssymb,amsthm}

\usepackage{graphicx}
\usepackage{tikz}
\usetikzlibrary{patterns,shapes,shapes.symbols,arrows,positioning}

\newtheorem{theorem}{Theorem}
\newtheorem{proposition}[theorem]{Proposition}

\newtheorem{definition}{Definition}
\newtheorem{remark}{Remark}

\DeclareMathOperator*{\argmax}{arg\,max}

\DeclareMathOperator{\E}{\mathbb{E}}

\title{Auditing Game-Theoretic Measures of\\Strategic Reasoning in LLMs}

\author{%
  Mateo Pechon-Elkins \\
  Yale University \\
  \texttt{mateo.pechon-elkins@yale.edu} \\
  \And
  Jon Chun \\
  Kenyon College \\
  \texttt{chunj@kenyon.edu} \\
}

\begin{document}

\maketitle

\begin{abstract}
Game-theoretic benchmarks can separate distinct forms of strategic behavior in LLMs that aggregate theory of mind scores do not distinguish, but only if those benchmarks measure what they claim. We audit a four-game framework and a sealed-bid auction control based on 1{,}855 interactions among seven LLMs, correcting its first release against the raw transcripts. A 1{,}024-token completion limit caused some models to return empty answers that parsers silently replaced with fixed actions, invalidating the reported Kimi K2 profile and contaminating measurements in all game types. A solution of the implemented signaling game also replaces the previously reported equilibrium bluff target of 0.340 with a conditional rate of $2/3$. On valid responses, bluff propensity ranges from $0.109$ to $0.732$ across models. A fitted model-specific propensity predicts held-out choices as well as or better than every equilibrium and QRE candidate tested. The equilibrium's type-level prescriptions are rejected for every model where they can be tested, and its aggregate rate keeps pace only when a model's own rate happens to be similar. Surviving valid cells cannot identify the previously reported cross-axis correlation. Finally, correcting an analysis-layer bug in the prompt-framing study reverses a prior result: reframing leaves Claude Haiku's bluff rate unchanged near $0.59$ but raises GPT-4o-mini's greatly, from $0.08$ to $0.65$--$0.67$. Framing effects are model-dependent, and the measurements remain framing-conditional. We therefore treat fitted $\lambda$ values as payoff- and model-dependent summaries instead of structural rationality parameters. The transcript corpus and per-move validity masks are available at \url{https://doi.org/10.5281/zenodo.21943627}. The exact solver and correction pipeline will accompany a future release. Appendix~\ref{sec:changelog} itemizes the changes.
\end{abstract}

\section{Introduction}

Theory of Mind (ToM), the ability to attribute mental states to others and use such attributions in predicting behavior, is vital to strategic interaction \citep{premack1978,wellman2014}. Originally characterized through false-belief paradigms in developmental psychology \citep{wimmer1983,baroncohen1985}, ToM is central to cooperation, deception, and multi-agent coordination \citep{apperly2009,wellman2014,schelling1960}.

As LLMs are increasingly used in negotiation, competitive games, and other strategic settings, evaluating ToM-relevant reasoning requires tools beyond aggregate ToM scores. Vignette-based tasks like Sally-Anne \citep{kosinski2023} are vulnerable to training-data contamination and often evaluate recognition rather than the intended strategic interaction ability \citep{ullman2023,shapira2023}; aggregate ``ToM scores'' conflate distinct capabilities \citep{chen2024,kim2023}. Without a formal grounding, high scores can reflect surface heuristics instead of belief modeling.

We address this gap by combining game-theoretic characterizations of four strategic games, including an exact equilibrium solve of the signaling game used (Section~\ref{sec:theory}); QRE-based bounded-rationality estimation with $\hat{\lambda}$ as the explicitly reduced-form fit statistic (Section~\ref{sec:qre}); and a Hoeffding-type finite-sample bound on aggregate win-rate concentration under Bradley--Terry (Section~\ref{sec:convergence}). Across 1{,}855 games with seven frontier LLMs (as well as four expansion models), response validity and prompt framing are the primary measurement-validity findings; Section~\ref{sec:results} reports the corrected numbers.

The first release of this preprint (v1, arXiv:2603.10029) stated several such results informally \citep{pechonelkinschun2026qre}; this version is a self-issued correction, formalizing the behavioral analysis (Section~\ref{sec:theory}), stating the finite-sample win-rate bound explicitly (Section~\ref{sec:convergence}), and reporting the corrected heterogeneity finding (Section~\ref{sec:results}).

Following behavioral game theory's as-if tradition of modeling strategic reasoning about others without the assumption of mutual belief consistency \citep{camerer2004}, we adopt a functional definition of ToM (Definition~\ref{def:tom}): an agent exhibits functional ToM if its behavior is consistent with both maintaining and updating models of others' mental states. Such a mechanism-agnostic definition enables measurement without entering debates about machine cognition. We make no claim that LLMs possess phenomenological ToM, nor that game performance generalizes to unstructured social interaction; the presented claims are restricted to behavioral signatures in structured game contexts.

\section{Related Work}

The study of ToM originated with \citeauthor{premack1978}'s \citeyearpar{premack1978} investigation of chimpanzee cognition and was later formalized using false-belief paradigms \citep{wimmer1983,baroncohen1985}. Developmental research established ToM as a multi-component capacity emerging between ages 3-5 \citep{wellman2014}, with distinct components of desire, belief, and attribution of knowledge-access acquired in ordered progression \citep{wellman2014}; \citet{apperly2009} further argue that belief tracking itself is served by two distinct systems.

Early LLM ToM evaluation approaches applied classic developmental tasks. \citet{kosinski2023} reported that GPT-4 solved 75\% of a battery of false-belief tasks, a rate the paper compared to six-year-old children; this result, though, was widely contested. \citet{ullman2023} demonstrated that minor task variations caused a collapse in performance, suggesting models often engaged in pattern matching rather than actual reasoning. \citet{shapira2023} systematically varied task parameters and found high brittleness, while \citet{sap2022} documented systematic failures in social reasoning requiring mental state attribution.

ToMBench \citep{chen2024} expanded the scope of evaluation with various scenarios. FANToM \citep{kim2023} stress-tests theory of mind in information-asymmetric conversations. These benchmarks evaluate comprehension of mental states, but they do not evaluate their use in strategic interaction. Our work requires dynamic, multi-round strategic interaction where success hinges on ongoing belief updating; formal equilibrium predictions then provide a validity check.

Recent work has highlighted the sensitivity of LLM evaluation to surface-level prompt variations. \citet{sclar2023} demonstrate that meaning-preserving changes in formatting (separator tokens, spacing, casing) can cause accuracy spreads of up to 76 points for a single model in few-shot settings, while \citet{mizrahi2024} show that paraphrased instruction templates change performance both in absolute terms and in relative model rankings. These findings raise concerns for any evaluation framework, including game-theoretic approaches: observed strategic behavior could reflect prompt-activated heuristics rather than underlying reasoning about payoff structure. We address this concern directly through prompt sensitivity analysis (Section~\ref{sec:promptsens}).

In game-theoretic LLM evaluation, SOTOPIA \citep{zhou2024} introduced LLM-vs-LLM social evaluation and automates scoring with GPT-4 (validated against human judgments), which may yet raise circularity concerns. \citet{gandhi2023} and \citet{lore2023} examined LLM behavior in specific games, showing that play depends strongly on prompt structure and context framing, with model-dependent sensitivity to game structure. Related work on deception and negotiation games generally reports aggregate outcomes without a formal decomposition.

Several recent benchmarks target LLM strategic reasoning. GTBench \citep{duan2024gtbench} utilizes 10 game tasks but does not estimate bounded-rationality parameters; WGSR-Bench \citep{wu2025wgsr} and LLMsPark \citep{chen2025llmspark} focus on aggregate performance rather than a parameter estimation with an accompanying uncertainty quantification. Most closely related, \citet{jia2025} apply truncated QRE (TQRE) to 22 LLMs in 13 single-shot abstract games. We differ from \citet{jia2025} in both scope and identification. We give a Hoeffding bound on aggregate pairwise win rates (they report TQRE fits without a finite-sample win-rate analysis) and estimate per-axis quantities rather than a single aggregate strategic score. Our multi-round games also require dynamic within-game belief updating, which cached single-shot heuristics cannot supply. After validity masking, this corpus cannot answer if the axes trade off (Section~\ref{sec:results}).

Quantal response equilibrium \citep{mckelvey1995} models bounded rationality by assuming agents choose actions with probabilities proportional to their expected utilities. Unlike Nash equilibrium, QRE accommodates systematic deviations from optimal play while still maintaining a game-theoretic structure. \citet{goeree2016} provide a detailed treatment of QRE theory and applications, including estimates of $\lambda$ for human subjects across a variety of games. We apply such a logit-QRE framework to LLMs, interpreting the fitted $\hat{\lambda}$ only as a reduced-form fit statistic: because $\lambda$ carries inverse-payoff units and is confounded with decoder temperature, it is not comparable across distinct payoff scales, and we make no comparison to human estimates (Section~\ref{sec:qre}).

The Bradley--Terry model \citep{bradley1952} provides probabilistic foundation for paired comparisons, and the Elo system operationalizes this through a system of sequential updates. Chatbot Arena estimates Bradley--Terry coefficients from large-scale pairwise LLM comparisons \citep{chiang2024}. We extend Elo with per-axis ratings and provide a Hoeffding-type concentration inequality on aggregated pairwise win rates, licensing the precision of the rating procedure's inputs.

\section{Game-Theoretic Framework}\label{sec:theory}

\begin{definition}[Functional ToM]
\label{def:tom}
An agent exhibits functional Theory of Mind when its behavior is consistent with: (a) holding models of other agents' epistemic and motivational states, (b) updating these state models from observed information, and (c) using such models to inform decisions.
\end{definition}

This definition is mechanism-agnostic. We decompose ToM-relevant capabilities into four separate axes, each measured by a game grounded in established theory (Table~\ref{tab:axes}, Figure~\ref{fig:games}).

\begin{table}[t]
\centering
\small
\begin{tabular}{@{}llll@{}}
\toprule
Axis & Capability & Game & Theory \\
\midrule
ESM & Epistemic State Modeling & Text-Dixit & Calibration \\
RSR & Recursive Strategic Reasoning & Strategic Claim & Signaling \\
SCG & Shared Conceptual Grounding & STST & Focal points \\
RSM & Relational State Modeling & Repeated PD & Cooperation theory \\
\bottomrule
\end{tabular}
\caption{Four-axis decomposition with theoretical foundations.}
\label{tab:axes}
\end{table}

This decomposition is motivated by developmental and cognitive accounts of ToM \citep{wellman2014,apperly2009}: ESM, RSR, SCG, and RSM target different behavioral capacities (epistemic state attribution, recursive belief modeling, mutual modeling for coordination, trust dynamics). Alternative, coarser decompositions exist (e.g., a belief/desire/intention grouping); the four-axis split is an operational design choice, and the corrected data does not adjudicate cross-axis structure (Table~\ref{tab:corr}).

\begin{figure}[t]
\centering
\begin{tikzpicture}[scale=0.48, every node/.style={font=\tiny}]
% Game 1: Strategic Claim (top-left)
\begin{scope}[xshift=0cm]
  \node[font=\scriptsize\bfseries] at (2.5,4.2) {Strategic Claim (RSR)};
  \draw[rounded corners, fill=blue!10] (0,0) rectangle (5,3.8);
  % Players
  \node[circle, draw, fill=white, minimum size=0.6cm] (p1) at (1,3) {P1};
  \node[circle, draw, fill=white, minimum size=0.6cm] (p2) at (4,3) {P2};
  % Private values
  \node[rectangle, draw, fill=yellow!30, minimum width=0.5cm] at (1,2.2) {v=3};
  \node[rectangle, draw, fill=yellow!30, minimum width=0.5cm] at (4,2.2) {v=?};
  % Claims
  \draw[->, thick] (1,1.8) -- (1,1.2) node[midway, right] {claim};
  \node[rectangle, draw, fill=green!20] at (1,0.8) {``5''};
  % Challenge
  \draw[->, thick, dashed] (4,1.8) -- (4,1.2) node[midway, left] {challenge?};
  \node at (2.5,0.3) {Bluff or honest?};
\end{scope}

% Game 2: Say the Same Thing (top-right)
\begin{scope}[xshift=6.5cm]
  \node[font=\scriptsize\bfseries] at (2.5,4.2) {Say the Same Thing (SCG)};
  \draw[rounded corners, fill=green!10] (0,0) rectangle (5,3.8);
  % Players
  \node[circle, draw, fill=white, minimum size=0.6cm] (p1) at (1,3) {P1};
  \node[circle, draw, fill=white, minimum size=0.6cm] (p2) at (4,3) {P2};
  % Words
  \node[rectangle, draw, fill=orange!30] at (1,2.2) {cat};
  \node[rectangle, draw, fill=orange!30] at (4,2.2) {dog};
  % Convergence arrows
  \draw[->, thick] (1.5,1.8) -- (2.2,1.2);
  \draw[->, thick] (3.5,1.8) -- (2.8,1.2);
  \node[rectangle, draw, fill=green!40] at (2.5,0.8) {pet?};
  \node at (2.5,0.3) {Find focal point};
\end{scope}

% Game 3: Repeated PD (bottom-left)
\begin{scope}[xshift=0cm, yshift=-5.2cm]
  \node[font=\scriptsize\bfseries] at (2.5,4.2) {Repeated PD (RSM)};
  \draw[rounded corners, fill=red!10] (0,0) rectangle (5,3.8);
  % Payoff matrix
  \draw (1,1.5) -- (4,1.5);
  \draw (2.5,0.5) -- (2.5,2.5);
  \node at (1.75,2.9) {C};
  \node at (3.25,2.9) {D};
  \node at (0.6,2) {C};
  \node at (0.6,1) {D};
  \node at (1.75,2) {3,3};
  \node at (3.25,2) {0,5};
  \node at (1.75,1) {5,0};
  \node at (3.25,1) {1,1};
  % Promise
  \node[cloud, draw, fill=white, cloud puffs=6, cloud puff arc=120, aspect=2, inner sep=0.5pt] at (4.2,2.8) {``I'll C''};
  \node at (2.5,0.3) {Trust \& cooperate?};
\end{scope}

% Game 4: Text-Dixit (bottom-right)
\begin{scope}[xshift=6.5cm, yshift=-5.2cm]
  \node[font=\scriptsize\bfseries] at (2.5,4.2) {Text-Dixit (ESM)};
  \draw[rounded corners, fill=purple!10] (0,0) rectangle (5,3.8);
  % Storyteller
  \node[circle, draw, fill=white, minimum size=0.5cm] (st) at (1,3) {ST};
  % Scenes
  \foreach \i in {0,1,2} {
    \draw[fill=gray!20] (2.5+\i*0.7,2.8) rectangle (3+\i*0.7,3.2);
  }
  \node[star, star points=5, draw, fill=yellow!50, minimum size=0.15cm] at (2.75,3) {};
  % Clue
  \draw[->, thick] (1.3,2.7) -- (2.3,2.3) node[midway, above] {clue};
  \node[rectangle, draw, fill=blue!20] at (2.5,1.8) {``floating''};
  % Guesser
  \node[circle, draw, fill=white, minimum size=0.5cm] (g) at (4,1.5) {G};
  \draw[->, dashed] (3.2,1.8) -- (3.7,1.6);
  % Confidence
  \node at (2.5,0.8) {Predict: 70\%};
  \node at (2.5,0.3) {Calibrate partner belief};
\end{scope}
\end{tikzpicture}
\caption{Schematic overview of the four games. Each targets a distinct ToM capability: Strategic Claim measures recursive reasoning through bluffing; STST measures conceptual grounding through word convergence; Repeated PD measures relational modeling through trust dynamics; Text-Dixit measures epistemic modeling through confidence calibration.}
\label{fig:games}
\end{figure}

Epistemic State Modeling (ESM) captures a capacity for modeling what another agent knows or believes, especially when this differs from one's own knowledge. Text-Dixit measures calibration accuracy in predicting how confidently a partner can identify a target from a clue. Recursive Strategic Reasoning (RSR) involves nested belief modeling. The Strategic Claim game is a Bayesian signaling game composed of payoff-relevant, challengeable messages (unlike the costless cheap talk of \citealp{crawford1982}), requiring calibrated deception and detection based on opponent modeling. Shared Conceptual Grounding (SCG) involves convergence on shared meaning through mutual state modeling. Say the Same Thing involves predicting focal points from mutual salience \citep{schelling1960}, with convergence driven by iterated reasoning about the partner's most likely choices \citep{nagel1995}. Relational State Modeling (RSM) tracks changing relationship dynamics. Repeated PD with Promises requires modeling both trust and commitment credibility over time, grounded in cooperation theory \citep{axelrod1981,embrey2018} and the folk theorem literature \citep{fudenberg1986}.

All games share the same design constraints. Conditions are procedurally generated, so memorization does not help; signals are observable, so beliefs can be updated; incentives create real strategic tradeoffs; and private information is sampled independently, making roles symmetric. Each game also admits an equilibrium characterization.

% Game-Theoretic Foundations (merged into Section 3; sec:theory label moved to section command above)

\subsection{Game 1: Strategic Claim (RSR Axis)}

Strategic Claim is a Bayesian signaling game $\Gamma_{SC}$ where each agent is given a private value $v \in \{1,...,6\}$ uniformly, claims a value $c \geq v$, and sets a challenge threshold $t \in \{1,...,7\}$. As implemented (and explained to the models), per-round payoff has three summed parts: a claim comparison (higher claim $+3$, tie $+1$, lower $0$), a challenger term (catching a bluff $+4$, false accusation $-2$), and a claimant term (caught bluffing $-4$, surviving a false challenge $+2$); a player challenges whenever the opponent's claim meets or exceeds their threshold (Definition~\ref{def:sc}, Appendix). \emph{Correction (v2):} the v1 manuscript's formal specification assigned different payoffs (claim value paid out, bluff bonus $c+2$, catch reward $+3$) and derived a behavioral target $\beta^* = 0.340$ from that stylized model; this stylized model does not describe the implemented game.

\begin{proposition}[Exact Stage-Game Equilibrium]
\label{thm:msne}
The implemented stage game $\Gamma_{SC}$ admits a symmetric mixed-strategy Bayesian Nash equilibrium computed by an exact sequence-form LCP solve of the Harsanyi extensive form (Gambit; exploitability zero to machine precision):
\begin{enumerate}[leftmargin=*,itemsep=1pt]
\item \textbf{Claims:} $v = 1$ mixes equally across claims $\{5, 6\}$; $v = 2$ mixes equally across claims $\{3, 4\}$; $v \geq 3$ claims honestly.
\item \textbf{Thresholds:} mix over $t \in \{4, 5, 6, 7\}$ with probabilities $(3/16,\, 3/16,\, 3/16,\, 7/16)$.
\end{enumerate}
At this equilibrium each challengeable claim level has posterior $P(\text{bluff} \mid c) = 1/3$, the challenger-indifference point in the implemented $+4/-2$ challenge payoffs. The conditional bluff rate given $v \leq 3$ is $\beta_{\mathrm{MSNE}} = 2/3$, and the overall bluff rate is $1/3$. No claim is made on equilibrium uniqueness. The v1 reference rate $\beta^* = 0.340$ is retained in this paper as a labeled historical reference; it is not a solution object of the implemented game: the v1 candidate profile that generated it is exploitable for $\approx 1.05$ points per round (Appendix~\ref{sec:proofs}).
\end{proposition}

This section does not put forth new equilibrium theory. What it adds is a unified treatment in which a single $\lambda$ parameterizes bounded rationality across four behavioral axes, the finite-sample win-rate bound that scopes per-axis comparisons (Section~\ref{sec:convergence}), and the corrected empirical picture of Section~\ref{sec:results}: LLM play is heterogeneous across models, and the dynamics are measured against the exact equilibrium above instead of a stylized target.

\subsection{Game 2: Repeated Prisoner's Dilemma (RSM Axis)}

The Repeated PD uses stage payoffs $(T,R,P,S) = (5,3,1,0)$ along with natural language cheap talk \citep{farrell1996} (formal specification in Definition~\ref{def:rpd}, Appendix). The v1 manuscript described a hidden horizon drawn from $\text{Uniform}\{7,15\}$ with a given known upper bound. As implemented, every game ran a \emph{fixed} 15-round horizon that was hidden from the players: a prompt displays the current round number but communicates no horizon information at all.

\begin{proposition}[Cooperation as Behavioral Finding]
\label{thm:folk}
For the \emph{full-information} finitely repeated PD with commonly known final round $T_{\max} = 15$, the unique subgame-perfect equilibrium (SPE) prediction is mutual defection in all rounds, from backward induction from $t = T_{\max}$. As played, however, such a horizon was not communicated: the game models faced is best modeled as a repeated PD with unknown horizon, and under sufficiently optimistic continuation beliefs, conditional cooperation is equilibrium behavior in such games \citep{fudenberg1986}. The observed sustained cooperation (valid-response block means 0.89--0.95 through round 15; Section~\ref{sec:results}) is thus reported as a behavioral finding consistent with the high early-round cooperation rates documented in repeated-PD trials \citep{embrey2018}, whose known-horizon settings still exhibit late-round unraveling---\emph{not} as a departure from the SPE of a known-horizon game. \emph{Correction (v2):} the v1 statement scored cooperation against the backward-induction benchmark of a known-horizon game; such a benchmark does not apply to this game as presented. The game remains informative for relational state modeling: sustaining cooperation requires modeling an opponent's willingness to reciprocate.
\end{proposition}

\subsection{Game 3: Say the Same Thing (SCG Axis)}

Say the Same Thing is a pure coordination game wherein players begin with distinct words and simultaneously select new words each round, aiming for exact match in 20 rounds or fewer. As implemented, a converged game scores $100 + 5 \cdot (20 - t)$ for \emph{both} players (identically), and a non-converged game scores $\max(0, (1 - d) \cdot 50)$ where $d$ is a word-distance heuristic; the v1 description ($1 - t/20$) was not correct. Under Schelling's focal point theory \citep{schelling1960}, efficient play requires mutual salience estimation. Because both agents always receive identical scores, every STST game is a rating draw and the SCG axis cannot differentiate models \emph{by construction}; the v2 diagnosis (Appendix~\ref{sec:stst_diag}) also found that 35.7\% of recorded STST moves are merely parser artifacts from truncated reasoning-model responses. The SCG axis is thus retired from all rating analyses.

\subsection{Game 4: Text-Dixit (ESM Axis)}

Text-Dixit is a signaling game where a storyteller agent observes a target among 6 surreal scenes, provides a clue, and then predicts the guesser's confidence. The guesser must select a scene and report confidence $p$. As implemented, game points are: $+2$ to a guesser for a proper identification, and $+1$ \emph{to the storyteller only} for a match in confidence calibration (prediction within 10 points of the guesser's reported confidence); the in-game message informing players that both earn a calibration point is incorrect, a discrepancy disclosed in v2. The analysis-level ESM metric also scores calibration by a quadratic loss, $100(1 - (\hat{p}/100 - p/100)^2)$, averaged at game level. Using a quadratic analysis metric \citep{gneiting2007}, a storyteller's optimal prediction is equal to the guesser's expected reported confidence, thereby requiring accurate epistemic state modeling of a partner's process of inference; the implemented $\pm 10$-point band payoff has a separate optimum, the windowed mode (Remark~\ref{prop:calib}, Appendix).

\subsection{Bounded Rationality: Quantal Response Equilibrium}
\label{sec:qre}

Real agents deviate systematically from Nash equilibrium. We formalize this concept via QRE \citep{mckelvey1995}.

\begin{definition}[Logit QRE]
\label{def:qre}
Given an expected utility $U_i(a_i, \sigma_{-i})$, the logit quantal response is:
\begin{equation}
\sigma_i(a_i | \lambda) = \frac{\exp(\lambda \cdot U_i(a_i, \sigma_{-i}))}{\sum_{a' \in A_i} \exp(\lambda \cdot U_i(a', \sigma_{-i}))}
\end{equation}
where $\lambda \geq 0$ is the rationality parameter. A QRE is a fixed point of this mapping.
\end{definition}

QRE exists for all $\lambda \geq 0$ (Brouwer fixed-point), ranges from uniform random ($\lambda \to 0$) toward play under Nash as $\lambda \to \infty$ (limit points are equilibria), and the logit specification is point-identified using choice data when actions have distinct expected payoffs \citep{mckelvey1995,goeree2016}. The selected games satisfy these identifiability conditions (Remark~\ref{thm:qre} and verification in Appendix).

We estimate $\lambda$ through maximum likelihood on per-round action data, treating each round as an independent stage-game draw:
\begin{equation}
\hat{\lambda} = \argmax_\lambda \sum_{i,t} \log \sigma_i(a_{i,t} | \lambda, \hat{\sigma}_{-i})
\end{equation}
where the optimizer is a bounded one-dimensional grid search with local refinement (the solver code, which will be included in a future release, cross-checks Newton-type restarts against the grid). To make the estimator reproducible from this text, we now state $\hat{\sigma}_{-i}$ and $U_i$ exactly as implemented.

\textbf{Opponent-behavior input $\hat{\sigma}_{-i}$.} For Strategic Claim, the opponent model is a scalar challenge-rate belief $\gamma_t$, initialized $\gamma_1 = 0.4$ and updated between rounds by exponential smoothing,
\begin{equation}
\gamma_{t+1} = 0.6\,\gamma_t + 0.4\,g_t, \qquad g_t = \begin{cases} 0.6 & \text{opponent's threshold} \leq 5\\ 0.2 & \text{otherwise;} \end{cases}
\end{equation}
for Repeated PD the model is a cooperation-rate belief $\kappa_t$, initialized $\kappa_1 = 0.5$ with $\kappa_{t+1} = 0.7\,\kappa_t + 0.3 \cdot \mathbf{1}[\text{opponent cooperated}]$. These are not fitted quantities, but fixed heuristics.

\textbf{Utility index $U_i$.} For Strategic Claim, the utility of a claim $c$ given a value $v$ is $U(c) = (1-\gamma_t)(c + 2) + \gamma_t(-4)$ for a bluff ($c > v$) and $U(v) = v + 0.3\,\gamma_t \cdot 2$ for the honest claim. \emph{Disclosure (v2):} this utility index retains the prior v1 stylized payoff model---it pays out the claimed value and a $+2$ bluff bonus, though the implemented game does not (Definition~\ref{def:sc}). For Repeated PD, $U(a)$ is the immediate expected stage payoff against $\kappa_t$ in addition to a damped continuation term, $U(a) = u(a, \kappa_t) + \min(1, r_t/10) \cdot 0.1 \cdot R_\delta(a)$, where $r_t$ is the number of rounds remaining and $R_\delta(a)$ is a geometric sum at $\delta = 0.91$ of the mutual-cooperation (respectively mutual-defection) stage payoff. The constants $0.4, 0.6, 0.2, 0.5, 0.7, 0.3, 0.91, 0.1$ are design choices. Thus, $\hat{\lambda}$ is a reduced-form fit statistic in a strong sense: it captures how sharply a model's choice frequencies react to a \emph{fixed heuristic utility index}, under i.i.d.\ assumptions that within-game learning violates. It is not equivalent to a structural rationality parameter of the implemented game, and so we do not compare it to human estimates. A v1 Bayesian layer (Gamma(2,1) prior over $\lambda$, posterior means and 95\% highest density intervals via grid approximation \citep{gelman2013}) was calculated on the v1 action set and is superseded by cluster-bootstrap intervals of Section~\ref{sec:results}; because any prior with mass away from zero pulls boundary estimates upward, boundary cells are reported as one-sided upper bounds. The prior-sensitivity analysis (Table~\ref{tab:prior_sens}, Appendix) showed posterior-mean variation of at most $0.05$ across Gamma(1, 0.5), Gamma(2, 1), and Gamma(3, 1) priors.

Because $\lambda$ multiplies payoffs inside the logit, it has inverse-payoff units: rescaling all payoffs by $s$ rescales the fitted $\lambda$ by $1/s$. We thus also report the payoff-normalized $\tilde{\lambda} = \hat{\lambda} \cdot \overline{\Delta U}$, where $\overline{\Delta U}$ is the mean per-decision utility range under the model's observation set. Finally, all data were collected at decoder temperature $T = 1.0$. As sampling temperature and $\lambda$ are both softmax inverse-temperatures, the fitted values confound model preferences with provider decoding and should therefore be read as $\lambda$-at-nominal-$T{=}1$ (Appendix temperature ablation; a token-logit experiment would be necessary to separate them).

\subsection{Convergence Analysis}
\label{sec:convergence}

The i.i.d.\ assumption used in the following theorems applies to \emph{game-level} outcomes: each complete game produces an independent result, despite the fact that within-game turns are non-stationary. Wald--Wolfowitz runs tests on game-level win/loss sequences across all involved 28 model pairings do not show evidence of serial dependence (all $p > 0.05$; see Remark~\ref{rem:iid}, Appendix).

\begin{proposition}[Pairwise Win-Rate Hoeffding Bound]
\label{thm:sample}
For models $i, j$ with a true win probability $p^*$ under the Bradley--Terry model, the empirical win rate $\hat{p}_n = \frac{1}{n}\sum_{t=1}^{n} S_t$ from $n$ i.i.d.\ games satisfies $P(|\hat{p}_n - p^*| > \epsilon) \leq 2\exp(-2n\epsilon^2)$. A 50-point Elo disparity corresponds to $p^* \approx 0.572$, requiring detection of $\epsilon = 0.072$, which gives $n \geq 356$ for $\alpha = 0.05$. We do not detect these differences from a single pairing's win rate; rather, we aggregate across all 28 pairings (1{,}855 games total, 100+ per model) and rely instead on a non-parametric bootstrap (median SD 40 Elo points). The bound therefore certifies the precision of \emph{aggregated} win rates instead of individual pairwise comparisons. Proof in Appendix~\ref{sec:proofs}.
\end{proposition}

\begin{proposition}[Within-Game Convergence in Expectation]
\label{thm:ingame}
If agents' effective behavior is consistent with exponential-smoothing belief updates with learning rate $\eta \in (0,1)$ and best-response slope $\kappa \in (-1,1)$ (such that $|\rho| < 1$ below), then the \emph{expected} conditional bluff rate $\beta_t$ (given $v \leq 3$) in Strategic Claim satisfies:
\begin{equation}
|\E[\beta_t \mid \beta_1] - \beta^\infty| \leq |\beta_1 - \beta^\infty| \cdot |\rho|^{t-1}, \quad \rho = 1 - \eta(1-\kappa)
\end{equation}
where $\beta^\infty$ is the fixed point of the belief--response map, treated as a free parameter estimated from data. \emph{Correction (v2):} the statement in v1 anchored this bound at stylized reference $0.340$; no equilibrium interpretation of $\beta^\infty$ is claimed. The bound is on conditional expectation of $\beta_t$; the realized bluff rate is stochastic and may deviate from this path in individual instances of games. We do not claim that LLMs literally implement exponential smoothing; it is a tractable illustrative smoothing model, and the pooled v2 series is non-monotone, so it is fit per model only (Figure~\ref{fig:permodel}, Appendix). Full derivation and assumptions in Appendix~\ref{sec:proofs}.
\end{proposition}

\section{Methodology}

\subsection{Models}

We evaluate seven frontier LLMs across four providers: GPT-5-mini and GPT-4o-mini\footnote{GPT-4o-mini was retired by OpenAI after data collection; replication would require substitution of a comparable model.} (OpenAI), Claude Haiku 4.5 (Anthropic), DeepSeek V3 (DeepSeek), Kimi K2 Thinking (Moonshot), and Gemini 2.5/2.0 Flash\footnote{Gemini 2.0 Flash has been deprecated by Google (shutdown March 2026).} (Google). This ensemble includes reasoning, efficient, and open-weight architectures (full specifications in Table~\ref{tab:models}, Appendix). All evaluations use a temperature $T=1.0$ such that mixed-strategy play, necessary for game-theoretic equilibria, is not suppressed.

\subsection{Experimental Design}
\label{sec:design}

From these seven models we construct 28 pairings (21 cross-model and 7 self-play); each pairing plays 10--18 replications per game type with pre-sampled conditions to ensure identical setups across comparisons. The full experiment comprises 1{,}855 games across five game types (Strategic Claim, Repeated PD, STST, Text-Dixit, and a sealed-bid first-price auction control providing a strategic-but-non-ToM baseline). Pre-sampled condition sets (150 per game type, e.g.\ STST starting words, SC value sequences, and Text-Dixit scenes) allow reproducibility and enable use of paired statistical tests. Ratings are batch Bradley--Terry refits of game-level outcomes with self-play excluded (the v1 online-Elo procedure with $R_a = 1500$ initialization, $K = 32$ update, and randomized game order was order-dependent and is superseded; Section~\ref{sec:results}), with non-parametric bootstrap CIs (1{,}000 resamples; bootstrap SDs 9--80 points, median 40). Proposition~\ref{thm:sample} gives a Hoeffding bound of $n \geq 356$ games per pairing for detecting a 50-point Elo difference from only win rates; the 10--18 replications per pairing achieve model-level precision by aggregation across 28 pairings (100+ games per model). QRE $\hat{\lambda}$ is estimated by bounded grid MLE on $[0,5]$ with local refinement (Newton-type restarts serve as a cross-check). Uncertainty is whole-game cluster-bootstrap percentiles, plus one-sided profile-likelihood upper bounds at boundary estimates ($\hat{\lambda} = 0$) where asymptotic normality fails (Section~\ref{sec:results}). \textbf{Compute resources:} all evaluations are frontier-LLM API calls; no local GPU is used. Total wall-clock $\approx$ 60 hours across 1{,}855 games + ablations, collected 13--15 February 2026; aggregate per-provider API cost was approximately \$80 (rates at collection). Reproducing the headline single-model $\hat{\lambda}$ fit takes minutes on a standard laptop.

\textbf{Data validity (v2).} All runs utilized a 1{,}024-token completion budget, and this is the corpus's central defect in data. Reasoning-class models frequently exhausted this budget on concealed reasoning, returning either empty or truncated visible responses that each game parser handled \emph{silently}: SC substituted (claim $= 1$, threshold $= 5$) in 99.5\% of Kimi K2's and 86\% of GPT-5-mini's SC rows; RPD's fallback records \emph{defection} for an empty response, affecting 78\% of Kimi K2's RPD rows; STST recorded regex artifacts as moves (35.7\% of STST moves; Appendix~\ref{sec:stst_diag}); the auction substituted the equilibrium bid $v/2$ (88.8\% of Kimi K2's rows). In Text-Dixit, 86\% of Kimi K2's acting moves fail the validity mask (85\% returned empty), so do 65\% of Gemini 2.5's (none empty: truncated outputs which had recorded actions overwritten with constants) and 24\% of GPT-5-mini's (empty responses). Gemini 2.5's SC and RPD rows are non-empty but truncated, with actions recovered using regex from its reasoning text; in Text-Dixit, where recovery often fails, 62.5\% of its moves were overwritten with constants (a further 12.5\% were truncated but still recorded faithfully). The other four models' rows are complete in SC, RPD, STST, and Text-Dixit; in the auction, a second backup channel (a bare-number regex that fires whenever JSON extraction fails) overwrote 1--11\% of Claude Haiku's, GPT-4o-mini's, and Gemini 2.0's recorded bids, which the validity mask is able to catch. Corpus-wide, 29.7\% of SC rows and 12.6\% of RPD rows are parser substitutions. Outcome records do not distinguish defaults from choices, so each per-model analysis in this paper is computed against a per-move validity mask, reported as ``valid responses'' alongside ``as recorded'' where they diverge, and claims that rested on parser-dominated cells, most prominently Kimi K2's whole strategic profile, are withdrawn (Section~\ref{sec:results}). A move is valid iff the recorded action matches the action that can be recovered from the raw response; an empty raw response always fails, and a non-empty truncated response whose recorded action was overwritten fails too (per-cell shares in Table~\ref{tab:validity}, Appendix).

\section{Results}
\label{sec:results}

\subsection{Behavioral Dynamics}

\begin{figure}[t]
\centering
\includegraphics[width=\textwidth]{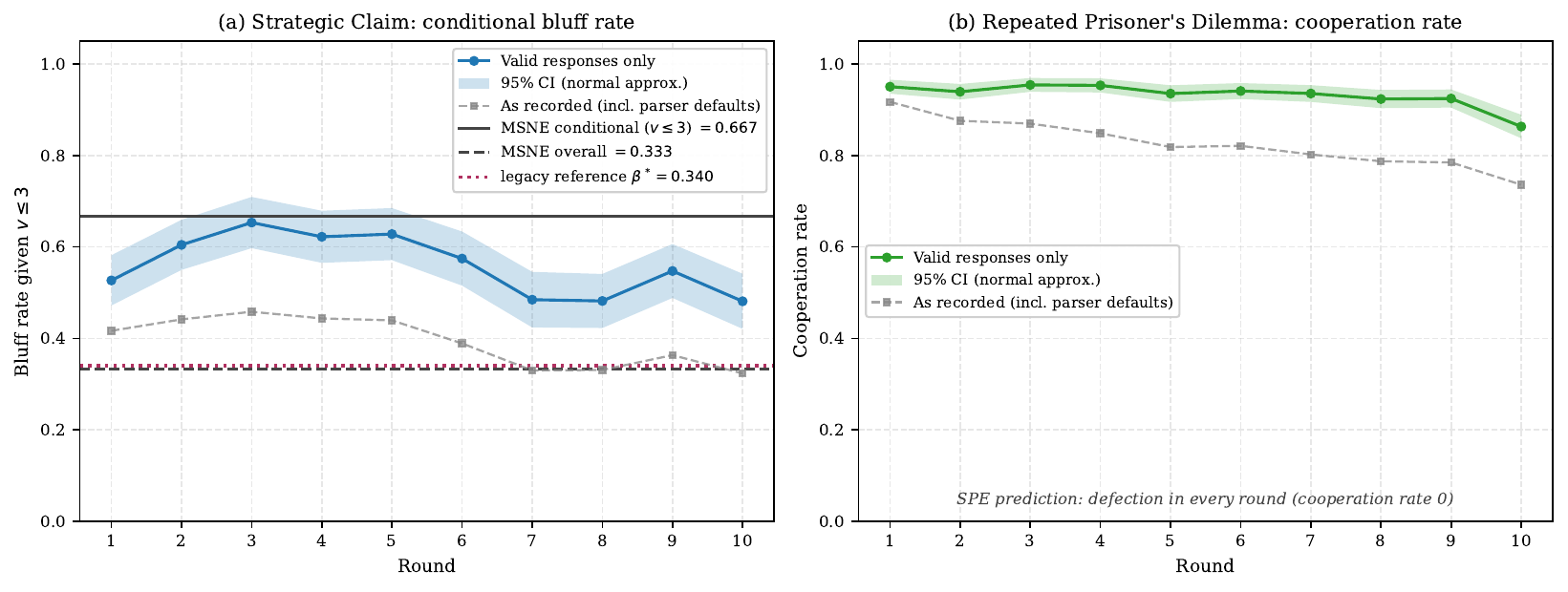}
\caption{Round-by-round behavioral dynamics (both players, 400 SC and 400 RPD games). Solid curves with 95\% CI bands: \emph{valid responses only}; faint dashed curves: \emph{as recorded}, including the parser defaults substituted for empty responses (claim $=1$/never-bluff in SC, defection in RPD). (a) Strategic Claim conditional bluff rate (given $v \leq 3$): valid-response bluffing runs $0.48$--$0.64$ and approaches none of the reference lines (exact-MSNE conditional $2/3$, MSNE overall $1/3$, legacy v1 reference $0.340$); the as-recorded curve's apparent settling near $0.33$--$0.34$ is default dilution, which is what v1 reported as convergence to target. (b) Repeated PD cooperation: valid-response cooperation stays at $0.89$--$0.95$ through round 15; the as-recorded decline reflects forced-defection defaults. The horizon was fixed and hidden, so no equilibrium benchmark applies (Proposition~\ref{thm:folk}).}
\label{fig:convergence}
\end{figure}

Figure~\ref{fig:convergence} shows round-by-round cooperation and bluff dynamics. Two layers of correction apply (change log, Section~\ref{sec:changelog}). First, the v1 series was computed on a filtered subset and is not reproduced by the raw records. Second, the outcome records themselves mix real model choices with \emph{parser defaults}: under the 1{,}024-token completion budget, Kimi K2 returned empty SC responses in 1{,}274 of 1{,}280 rows and GPT-5-mini in 1{,}101 of 1{,}280, each silently recorded as a fallback action (claim $= 1$, threshold $= 5$); in RPD, the fallback records \emph{defection} for an empty response, affecting 1{,}498 of Kimi K2's 1{,}920 rows. All other models returned non-empty responses in every SC row; in RPD, GPT-5-mini returned 19 empty rows (1.0\%) and the remaining models none. We thus report the valid-response series (rows whose raw response is non-empty) as primary, with as-recorded series shown for contrast.

Among valid responses, the aggregated conditional bluff rate runs high and non-monotone---$0.57 \pm 0.02$, $0.64 \pm 0.02$, $0.60 \pm 0.02$, $0.48 \pm 0.02$, $0.52 \pm 0.02$ across round blocks (Table~\ref{tab:convergence}, Appendix)---approaching \emph{none} of the reference values: it sits between the v1 stylized reference ($0.340$) and the exact-MSNE conditional rate ($2/3$), though it is closer to the latter in earlier rounds. The as-recorded series (including parser defaults) instead drifts to $0.344 \pm 0.017$ by rounds 9--10, simultaneously numerically near the v1 target and the equilibrium's \emph{overall} bluff rate ($1/3$). The v1 ``convergence to within 4\% of target'' claim thus failed on two counts: the actual target was a conditional/overall confusion (Proposition~\ref{thm:msne}), and the apparent approach to $0.34$ was in large part produced by never-bluffing parser defaults diluting the pool and contaminating accuracy. We report these dynamics descriptively and do not make any equilibrium-convergence claim. Cooperation in RPD among valid responses is high and stable---$0.945 \pm 0.006$ (rounds 1--2) to $0.894 \pm 0.008$ (rounds 9--10), still $0.895 \pm 0.005$ across rounds 11--15; the as-recorded decline to $0.76$ was largely Kimi K2's forced-defection defaults. Sustained cooperation is consistent with the high cooperation rates commonly documented in repeated-PD experiments \citep{embrey2018} and is reported without an equilibrium benchmark because the horizon was not available to players (Proposition~\ref{thm:folk}).

\subsection{QRE Rationality Parameters}

$\hat{\lambda}$ estimates with cluster-bootstrap uncertainty are reported in Table~\ref{tab:qre_cv} (Appendix). \emph{Correction (v2).} Two defects were compounded in the v1 table. First, the v1 estimator's action set $\{v, \dots, 6\}$ was unable to represent claims below value, so corresponding rows were silently dropped. Second, almost all rows with $c < v$ are not model choices at all: they are a parser default (claim $= 1$, threshold $= 5$) substituted for empty responses when the extended-reasoning models (Kimi K2 and GPT-5-mini) exhausted their 1{,}024-token budget. Kimi K2 has 1{,}274 default rows out of 1{,}280 (and \emph{zero} chosen under-claims); GPT-5-mini has 1{,}101 (zero chosen; three further rows match the default constant but are real truthful claims at $v = 1$); genuinely chosen under-claims by responding models are rare yet present (3.2\% for DeepSeek V3, $\leq 0.5\%$ elsewhere) despite being forbidden by the prompt and unenforced by the engine. Consequently, the v1 $\hat{\lambda}$ for Kimi K2 and GPT-5-mini characterized a parser constant, and no SC $\hat{\lambda}$ is estimable for those two models from this corpus (their cells are marked accordingly). For those five models whose SC rows are real responses, re-estimation on the full legal claim set with complete rows gives: GPT-4o-mini 1.11 [0.93, 1.31], Gemini 2.0 0.85 [0.73, 1.00], Gemini 2.5 0.77 [0.70, 0.86], DeepSeek V3 0.62 [0.56, 0.68], Claude Haiku 0.53 [0.47, 0.58]. In RPD, the prior Kimi K2 $\hat{\lambda} = 1.10$ is retracted as a parser artifact: 1{,}498 of its 1{,}920 RPD rows are empty responses that the fallback default recorded as \emph{defection}, and a $\lambda$ fitted to that series measures only the parser; the remaining models (all with complete real responses) rest at the boundary with one-sided 95\% bounds $\leq 0.002$. Intervals are whole-game cluster-bootstrap percentiles ($B = 2000$); boundary cells carry one-sided profile-likelihood bounds (Chernoff mixture cutoff), rather than suppressed standard errors. Even for estimable cells, the held-out comparison of Section~\ref{sec:modelcomp} shows a fixed bluff propensity outperforming QRE on the decision margin, so we refrain from interpreting $\hat{\lambda}$ as evidence of bounded-rational expected-utility responsiveness. Identifiability is outlined two ways in the Appendix: profile likelihoods per cell (Figure~\ref{fig:profilell}) and a parameter-recovery study (Figure~\ref{fig:recovery}). The v1 Bayesian layer (Gamma$(2,1)$ posteriors) inherited the truncated set of actions and is superseded by bootstrap intervals; its prior-sensitivity analysis is still retained for the record (Table~\ref{tab:prior_sens}, Appendix). Per-model convergence trajectories (Figure~\ref{fig:permodel}, Appendix) show heterogeneous dynamics.

\subsection{Per-Axis Bradley--Terry Ratings}

\begin{table}[htbp]
\centering
\small
\setlength{\tabcolsep}{5pt}
\begin{tabular}{@{}lcccc@{}}
\toprule
Model & ESM & RSR & RSM & Control \\
\midrule
Claude Haiku & 1398 (1399) & \textbf{1562} (1720) & 1480 (1340) & 1545 \\
DeepSeek V3 & 1355 (1418) & 1507 (1656) & 1494 (1364) & \textbf{1614} \\
Gemini 2.0 & 1415 (1466) & 1426 (1574) & 1446 (1313) & 1586 \\
Gemini 2.5 & --- & 1488 (1530) & \textbf{1582} (1450) & --- \\
GPT-4o-mini & 1329 (1412) & 1517 (1693) & 1426 (1282) & 1161 \\
GPT-5-mini & \textbf{2002} (1679) & --- & 1574 (1458) & 1593 \\
Kimi K2 & --- & --- & --- & --- \\
\bottomrule
\end{tabular}
\caption{Per-axis batch Bradley--Terry ratings, \emph{validity-masked} (as-recorded values shown in parentheses); bold marks axis leader among valid cells; ``---'' denotes model-axis cells dropped for response validity (with moves valid iff the recorded action matches the action recoverable from the raw response; a model keeps an axis only if $\geq 50\%$ of its acting moves are valid). Removed: Kimi K2 everywhere (0.5--22\% valid, dependent on axis); GPT-5-mini on RSR (14\%); Gemini 2.5 on ESM (35\%, truncation-overwritten Text-Dixit moves) and Control (no games). SCG omitted: STST scores are pinned at 1500 by construction (Appendix~\ref{sec:stst_diag}).}
\label{tab:elo}
\end{table}

Rank reversals persist in valid cells (Table~\ref{tab:elo}): Claude Haiku tops RSR but sits near the bottom of ESM, and GPT-5-mini leads ESM by a large margin but has no interpretable RSR rating. The masked ranges are, however, far more compressed than v1 reported (RSR spans 136 points across the five valid models, versus 626 as recorded), because much of the v1 spread \emph{was} due to the validity artifact. The mask matters through a \emph{zombie-opponent effect}: wins over models that were silently playing engine defaults unfairly inflated survivors' RSR ratings by 43--176 points (score rate 0.82--1.0 against dropped models vs.\ 0.35--0.66 against retained ones), and forced-defection defaults decreased survivors' RSM ratings by 116--144. v1's online Elo was additionally order-dependent and tainted by self-play updates; the v2 batch refit is order-independent and does not involve self-play. We refrain from any per-pairing significance claims: with 10--18 replications per pairing the Hoeffding bound (Proposition~\ref{thm:sample}) does not certify a per-pair detection of 50-point differences.

Cross-axis correlations necessitate three layers of correction. \emph{Layer 1 (estimator):} v1's $r = -0.95$ came from a corrupted order- and self-play-noisy online Elo; on batch BT ratings with a full-refit game bootstrap ($B = 5000$) and Holm-corrected exact permutation tests, the recorded values are ESM--RSR $-0.81$ (CI $[-0.93, -0.54]$, Holm $p = 0.046$), RSR--RSM $-0.82$ (CI $[-0.89, -0.70]$, Holm $p = 0.018$), ESM--RSM $+0.39$ (n.s.; in v1 $+0.72$ was estimator noise). The v1 ``$p = 0.0009$'' was a parametric $p$ improperly labeled as a permutation result. \emph{Layer 2 (disposition):} on as-recorded data the trade-off is not distillable to bluff propensity ($r(\text{ESM},\text{RSR} \mid \beta) = -0.80$, CI excluding zero).

\emph{Layer 3 (validity---new in v2, decisive):} the as-recorded correlations are calculated over model-axis cells that the validity audit makes invalid, and the contaminated models sit precisely where they drive correlation (Kimi K2 and GPT-5-mini are the RSR floor \emph{because} their SC moves were largely parser constants). Restricted to only valid cells, ESM--RSR has $n = 4$ models: $r = -0.36$, bootstrap CI $[-0.96, +0.85]$, and the exact permutation floor at $n = 4$ is $2/4! = 0.083$, so \emph{no} correlation on valid data could reach $p < 0.05$; RSR--RSM at $n = 5$ collapses to $+0.04$ (and is threshold-sensitive: $-0.39$ under a laxer within-game validity cutoff); ESM--RSM at $n = 5$ is $+0.89$ (floor-limited, Holm $p = 0.25$).

A decomposition pins down \emph{why} the trade-off collapses: computing ESM--RSR over the same four uncorrupted models but with their \emph{as-recorded} ratings gives $r = -0.98$, which falls to $-0.36$ only when the models' ratings are refit on valid games---i.e., the clean models' own ratings had been distorted by games played against opponents whose moves the engine was actually fabricating. We also note, without pressing it (it cannot clear a permutation floor either), that axes remaining jointly measurable correlate strongly \emph{positively} on the valid data ($+0.89$)---which would be the signature of a general competence factor rather than the dissociability v1 claimed. The conclusion supersedes v1 and the intermediate correction: this corpus cannot either establish or refute an empathic--adversarial trade-off. The v1 headline correlation is retracted as unanswerable from these data; proper settlement would require a corpus in which every model's responses survived validity auditing.

\emph{Retraction (v2): the discriminant-validity control.} The v1 manuscript reported a set of near-zero auction-control correlations ($+0.09$, $+0.11$, $+0.14$) as evidence that the axes could capture more than merely generic strategic competence. This row is not reproducible from transcripts under any construction we tried (every reconstruction of Control--RSR is instead \emph{negative}), the control covers only six models (Gemini 2.5 did not play auction games), and 88.8\% of Kimi K2's auction responses were not usable and silently replaced by a parser default---which is the equilibrium bid, making a non-responding model falsely look optimally calibrated. In the actual data, the control correlates with the axes at $|r| = 0.34$--$0.43$ (none significant at $n = 6$, none near zero). The discriminant-validity claim is withdrawn; the auction data cannot support it in either direction.

Table~\ref{tab:predictions} (Appendix) outlines the framework-level quantities under the v2 corrections: valid-response cooperation holds at 0.89--0.95 across round blocks (the as-recorded 76\% is diluted by the parser-forced defection defaults; no equilibrium benchmark can attach, since the horizon was hidden; Prop.~\ref{thm:folk}); aggregated ratings stabilize with a bootstrap SD 9--80 points (median 40; Section~\ref{sec:design}); valid-response late-round bluffing ($0.515 \pm 0.022$, rounds 9--10) is approximately three-quarters of the exact-MSNE conditional rate $2/3$ (Prop.~\ref{thm:msne}); and no $\lambda$--performance correlation is reported, because SC $\hat{\lambda}$ is only estimable for five models and the estimable RPD cells are merely boundary values, so that v1 correlation has no v2 counterpart (Table~\ref{tab:qre_cv}). On axis structure, one sentence replaces the contradictory v1 statements: \emph{rank reversals across axes persist among valid cells, but the cross-axis correlation structure, including the earlier v1 trade-off headline, is not answerable from this corpus after parser-contaminated cells are excluded.}

\subsection{Prompt Sensitivity as a Measurement-Validity Finding}
\label{sec:promptsens}

The largest robustness effect measured concerns prompt framing, and in this version it is reported as a primary finding rather than simply an appendix limitation, with the v1 numbers corrected. We ran SC self-play under three distinct prompt variants intended as semantically equivalent descriptions of an identical incentive structure: \emph{original} (game-theoretic narrative with ``claim'' and ``challenge'' language), \emph{formal} (academic mechanism description), and \emph{minimal} (bare payoff structure), for GPT-4o-mini and Claude Haiku, 10 games per condition (60 games).

\emph{Correction (v2).} The v1 table reported a $\beta = 0.000$ in all four reframed cells and read the result as the framing eliminating bluffing. Those zeros were, however, an analysis-layer artifact, not measured model behavior: the variant reframings omitted the round-history field from the logged model-input record, the bluff-rate, round-10, and $\lambda$ estimators read observations from that field, and an empty observation set was misreported as $0.0$ rather than as missing. Those transcripts are the cleanest in the corpus: 1{,}200 acting moves with no empty responses and no parse failures, and one overwritten move (Claude Haiku's original cell) under the per-move validity mask used throughout. Table~\ref{tab:prompt_sens} reports rates recomputed from the corrected raw outcome records.

\begin{table}[htbp]
\centering
\small
\begin{tabular}{@{}llcccc@{}}
\toprule
Model & Variant & Moves (valid) & $\beta$ ($\pm$SE) & $n_{v \leq 3}$ & $\beta_{R10}$ ($n$) \\
\midrule
GPT-4o-mini & original & 200 (200) & $0.079 \pm 0.027$ & 101 & 0.000 (11) \\
GPT-4o-mini & formal & 200 (200) & $0.674 \pm 0.049$ & 92 & 0.818 (11) \\
GPT-4o-mini & minimal & 200 (200) & $0.650 \pm 0.047$ & 103 & 0.556 (9) \\
\midrule
Claude Haiku & original & 200 (199) & $0.588 \pm 0.050$ & 97 & 0.222 (9) \\
Claude Haiku & formal & 200 (200) & $0.589 \pm 0.050$ & 95 & 0.400 (10) \\
Claude Haiku & minimal & 200 (200) & $0.577 \pm 0.050$ & 97 & 0.700 (10) \\
\bottomrule
\end{tabular}
\caption{Prompt sensitivity for SC self-play, recalculated from raw outcome records (v2 correction: the v1 zeros in reframed cells were an analysis-layer artifact). $\beta$ is bluff rate conditional on $v \leq 3$ over valid moves (10 games per condition; both seats); $\beta_{R10}$ is round-10 bluff rate with cell sizes of 9--11 observations, indicative only. The v1 $\hat{\lambda}$ column is dropped: the fits inherited the same defect, and the corrected cells do not meet this paper's uncertainty-reporting standards at current sample sizes. Reframing leaves Claude Haiku's bluff rate essentially unchanged and raises GPT-4o-mini's by roughly a factor of eight.}
\label{tab:prompt_sens}
\end{table}

The corrected result is now a model-dependent framing effect: Claude Haiku's conditional bluff rate is essentially constant across framings ($0.588$, $0.589$, $0.577$), while GPT-4o-mini's rises by roughly a factor of eight once the narrative is stripped ($0.079$ original; $0.674$ formal, $0.650$ minimal). Two wording confounds qualify the mechanistic reading: only the original variant uses the word ``bluff'' (with a worked bluffing example), and the reframed variants actually display the legal claim range as the live interval $[v, 6]$, making the room above one's value more salient each round, whereas the original states the constant range 1--6 with the floor in prose only; separating narrative framing from instruction salience requires a factorial design we have not run. A measurement-validity conclusion persists in sharpened form: framing effects are both large and heterogeneous across models \citep{ullman2023,sclar2023,mizrahi2024}, so measured bluff propensities are framing-conditional, and cross-model comparisons are internally consistent (all models saw identical prompts in the main corpus) but not framing-invariant. That the v1 zeros were themselves a silent analysis-layer default is also an instance of the pathology this audit documents, caught by validating the estimator against the raw transcripts. With 10 self-play games per condition these estimates can be used to bound the direction of the framing effect; a powered, preregistered, cross-play framing study is the most important follow-up such a framework needs.

\section{Discussion}

We set out a methodology designed for estimating bounded rationality in AI agents under strategic uncertainty, and in v2 a worked example of what such estimates require: an exact equilibrium solve of the implemented game, a common event space for model comparison, full-refit uncertainty on rating-derived correlations, and a parameter-recovery study that bounds where the estimator is informative. QRE parameter inference here returns reduced-form, payoff-scale-dependent fit statistics; we make no claim of any kind about calibration against human behavioral data.

Kimi K2's earlier v1 profile (RSM dominance, ``strategic defection timing,'' the RPD $\hat{\lambda} = 1.10$) is primarily a token-budget artifact: 78\% of its RPD responses and 99.5\% of its SC responses were empty (instances where hidden reasoning exhausted the 1{,}024-token budget), and the parsers silently recorded defection and claim-$1$ defaults (Figure~\ref{fig:rpd_cooperation}; Appendix~\ref{sec:changelog}). A ``model'' that defects when it fails to answer dominates six cooperative opponents (the surviving models' score rate against it was exactly $0.000$) and looks strategically sophisticated in the way v1 reported. The same applies to GPT-5-mini's SC data (86\% defaults). Among models with genuine responses, the dissociation between precisely \emph{how often} and \emph{how sharply} choices actually track the utility index persists: GPT-4o-mini bluffs least ($\beta = 0.109$) but has the highest SC $\hat{\lambda} = 1.11$. We caution against over-reading either quantity as definitive strategic capability, given the held-out result that a fixed propensity out-predicts QRE where the data are genuine (Table~\ref{tab:bic}). Response-validity auditing should thus precede behavioral interpretation for models where hidden reasoning can exhaust the completion budget (Methodology, Data validity). Model expansion (Table~\ref{tab:model_expansion}, Appendix) reveals non-monotonic $\hat{\lambda}$ trajectories across versions on the earlier v1 self-play pipeline, with the same validity caveat attached. The SCG axis (STST) is now retired: its zero rating variance is an artifact of identical seat scoring, and a full third of its recorded moves are parser artifacts (Appendix~\ref{sec:stst_diag}).

\paragraph{Limitations.} Prompt-framing sensitivity is reported as a central result (Section~\ref{sec:promptsens}) rather than delineated here; the remaining limitations are: (1) $n = 7$ model-level points do not support dimensionality claims, and all correlational structure is descriptive with dual (game- and model-sampling) uncertainty as reported in Section~\ref{sec:results}; (2) the $\lambda$/temperature confound is not resolved---all data were collected at $T = 1.0$, the ablation (Table~\ref{tab:temp_ablation}, Appendix) is underpowered at only 10 games per cell, and separation would require token-level logit access; (3) the 1{,}024-token completion budget corrupted STST and auction data for extended-reasoning models (Methodology, Data validity), and the affected claims are retracted rather than repaired; (4) observed dynamics are consistent with simple adaptive mechanisms, and causal claims about LLM opponent modeling would necessitate scripted-partner interventions, which we have left to future work; (5) models evolve rapidly, and the non-monotone version trajectories in the expansion study strongly caution against reading any snapshot as a static capability. We intend this as methodology for ongoing evaluation, not a fixed benchmark. Appendix~\ref{sec:changelog} records the full v1-to-v2 correction trail against which later re-runs can be checked.

\section{Conclusion}

We have put forth a game-theoretic evaluation methodology for axis-specific strategic behavior in LLMs---and, in this version, corrected its first release against raw response records. For game-based LLM evaluation, measurement validity is a harder problem than capability ranking. One benchmark produced all of the following: parser defaults recorded as choices, a wrong equilibrium target, a cross-axis correlation that turns out to be unanswerable, and a framing-collapse result fabricated by a silent analysis-layer zero. Reasoning-class models with hidden deliberation that may silently exhaust token budgets are especially exposed. Non-monotone version trajectories argue in favor of continuous re-evaluation, and two of the seven panel models are retired. The full 1{,}855-game transcript corpus and its per-move validity masks are available at \url{https://doi.org/10.5281/zenodo.21943627}. The exact solver and correction pipeline will accompany a future release.

\bibliography{neurips2026_gtom_arxiv_20260822}

\newpage

\appendix

\section{Changes from v1}
\label{sec:changelog}

This version is a self-issued correction of arXiv:2603.10029v1 (submitted February 2026, announced March 2026), which presented with the title ``Quantal Response Equilibrium as a Measure of Strategic Sophistication: Theory and Validation for LLM Evaluation.'' The changes below are detailed where they occur:

\begin{enumerate}[leftmargin=*,itemsep=1pt]
\item \textbf{Formal game specification corrected.} The v1 Definition of Strategic Claim stated payoffs that were not implemented; Definition~\ref{def:sc} now states the implemented payoffs, so the discrepancy is disclosed (Section~\ref{sec:theory}).
\item \textbf{An exact equilibrium solve replaces the stylized target.} The implemented game is now solved exactly (sequence-form LCP; Proposition~\ref{thm:msne}): the MSNE conditional bluff rate is $2/3$ rather than $0.340$. The v1 ``convergence to within 4\% of target'' claim had conflated the equilibrium's conditional and overall bluff rates and is now withdrawn; the dynamics are now reported descriptively (Section~\ref{sec:results}).
\item \textbf{Model-comparison table rebuilt on common event space.} The v1 BIC table evaluated candidates using model-dependent event spaces and row sets (this implied uniform action counts 2.9--5.9); on a common event space with complete rows, 4 of 7 best-fit labels are changed and the earlier v1 three-cluster claim is now retracted (Appendix, Table~\ref{tab:bic}).
\item \textbf{Human-$\lambda$ calibration removed.} $\hat{\lambda}$ is now reported only as a unit-caveated, reduced-form fit statistic $\tilde{\lambda}$, with the decoder-temperature confound explicitly stated (Section~\ref{sec:qre}). Any v1 human-range comparisons and elements of the human-baseline appendix are deleted.
\item \textbf{Estimator made reproducible; boundary cells receive one-sided bounds.} $\hat{\sigma}_{-i}$ and $U_i$ are now comprehensively specified (Section~\ref{sec:qre}). $\lambda$ tables use full-game cluster-bootstrap intervals, with at-boundary one-sided upper bounds; both profile likelihoods and a parameter-recovery study delimit identifiability (Appendix).
\item \textbf{Correlation inference repaired, withdrawn as unanswerable.} The full-refit game-level bootstrap (named but unrun in v1) is now run, with corresponding Holm-corrected exact permutation tests and a bluff-disposition partial-correlation control; estimator repair moves the earlier headline number from $-0.95$ to $-0.81$, and the validity mask shows that the remaining data ($n = 4$ valid model-axis cells for ESM--RSR) are unable to answer the cross-axis question. The contradiction on axis independence in the abstract/discussion is resolved by withdrawal (Section~\ref{sec:results}).
\item \textbf{Prompt-sensitivity numbers revised; result reversed.} The original v1 framing table falsely reported a $\beta = 0.000$ under the formal and minimal reframings; such zeros were actually an analysis-layer artifact (reframed variants omitted the round-history field read by estimators, and the empty observation set was reported as $0.0$). Recomputed from the raw outcome records instead (1{,}200 moves, 1{,}199 valid), reframing has Claude Haiku's bluff rate unchanged and magnifies GPT-4o-mini's by a factor of eight; the framing-collapse claim is withdrawn and succeeded by a model-dependent framing effect, moved to the main results (Section~\ref{sec:promptsens}).
\item \textbf{Numbers regenerated from raw records.} Some v1 statistics (for instance, round-block dynamics) were computed on filtered subsets and thus did not reproduce; all v2 numbers regenerate from the outcome records via the specified correction pipeline.
\item \textbf{Response-validity audit; Kimi K2 strategic profile retraction.} A forensic pass over raw responses discovered that the 1{,}024-token completion budget caused reasoning-class models to often return empty or truncated answers read by game parsers as defaults: claim $= 1$ in SC (99.5\% of Kimi K2's and 86\% of GPT-5-mini's rows), \emph{defection} in RPD (78\% of Kimi K2's rows), the equilibrium bid in the auction, and regex artifacts in STST. Kimi K2's original v1 profile (RSM dominance, ``strategic defection timing,'' RPD $\hat{\lambda} = 1.10$) is now retracted as merely a parser artifact; behavioral statistics are calculated and reported on valid responses throughout, with the as-recorded values shown for contrast (Methodology, Data validity; Section~\ref{sec:results}).
\end{enumerate}

\section{Formal Game Definitions and Additional Results}
\label{sec:formal_defs}

\begin{definition}[Strategic Claim Stage Game (with implemented payoffs)]
\label{def:sc}
The implemented Strategic Claim game is a Bayesian signaling game $\Gamma_{SC} = (N, V, A, u, p)$ where $N = \{1, 2\}$, $V = \{1,...,6\}$ with $p(v) = 1/6$, action space $A_i = C_i \times T_i$ with $C_i = \{v_i,...,6\}$ (claims $\geq$ value) and $T_i = \{1,...,7\}$ (possible challenge thresholds; player $i$ challenges iff $c_j \geq t_i$). Player $i$'s per-round payoff is the sum of three separate constituent parts: \textbf{claim comparison} ($+3$ if $c_i > c_j$; $+1$ if $c_i = c_j$; $0$ otherwise); \textbf{challenger term}, if $i$ challenges $j$ ($+4$ if $j$ bluffed, i.e.\ $c_j > v_j$; $-2$ otherwise); and \textbf{claimant term}, if $j$ challenges $i$ ($-4$ if $i$ bluffed; $+2$ otherwise). The claimed value does not directly pay out. \emph{Correction (v2):} the v1 manuscript stated different payoffs (honest claim $= c$, unchallenged bluff $= c+2$, catch reward $+3$); those payoffs were not implemented, and the v1 derivations built on them (in particular $\beta^* = 0.340$) are superseded by the new exact solve in Proposition~\ref{thm:msne}.
\end{definition}

\begin{definition}[Repeated PD with Cheap Talk (implemented)]
\label{def:rpd}
$\Gamma_{RPD} = (N, A, u, T, M)$ with stage payoffs $(T,R,P,S) = (5,3,1,0)$, a hidden, \emph{fixed} horizon of 15 rounds (the prompt only displays the current round number and no horizon information), and message space $M$ (natural language cheap talk is exchanged before each decision). \emph{Correction (v2):} the v1 definition stated a random horizon $\text{Uniform}\{7,15\}$ with commonly known upper bound; instead, all logged games ran 15 rounds, and no horizon information was communicated. Consequently, an SPE-defection benchmark of the known-horizon version of the game does not describe the game as presented (Proposition~\ref{thm:folk}).
\end{definition}

\begin{definition}[STST Coordination Game (implemented)]
Players begin with distinct words $(w_1^0, w_2^0)$. Each round $t$: players must simultaneously select words $w_i^t$; the game ends when $w_1^t = w_2^t$ (exact string match after normalization) or $t = 20$. Score, identical for both players: $100 + 5(20 - t)$ if converged; $\max(0, (1 - d) \cdot 50)$ otherwise. \emph{Corrections (v2):} the v1 payoff ($1 - t/20$) was incorrect; the distance $d$ is, as run, a character-set Jaccard fallback, not a documented sentence-embedding distance (the embedding dependency was absent); and identical seat scores mean the game carries no direct between-model rating information (Appendix~\ref{sec:stst_diag}).
\end{definition}

\begin{definition}[Text-Dixit Signaling Game]
Storyteller observes a target scene $s^*$ among 6 scenes $S = \{s_1,...,s_6\}$, selects clue $c$ from word bank, and predicts confidence of the guesser $\hat{p}$. Guesser observes $(c, S)$, selects their guess $\hat{s}$, reports confidence $p$. ESM score $= 100(1 - (\hat{p}/100 - p/100)^2)$.
\end{definition}

\begin{remark}[Calibration Targets]
\label{prop:calib}
Two distinct optimality targets are attached to the prediction of the storyteller. Under the \emph{implemented} game payoff (the $\pm 10$-point band: $+1$ iff $|\hat{p} - p| \leq 10$), the optimal prediction maximizes $P(|\hat{p} - p| \leq 10)$, a windowed mode of the confidence distribution of the guesser. Under the \emph{analysis-level} quadratic metric, the optimum is a conditional mean $\hat{p}^* = \E[p \mid c]$; the two coincide for only symmetric, unimodal confidence distributions, which confidence reports do not necessarily follow. If the guesser is well-calibrated, $\E[p \mid c] = P(\text{guesser correct} \mid c)$. Overconfidence suggests poor perspective-taking; underconfidence suggests too much uncertainty about the partner's inference.
\end{remark}

\begin{remark}[QRE Properties---cited background]
\label{thm:qre}
The logit QRE satisfies: (1) \textbf{existence} for all $\lambda \geq 0$ (Brouwer fixed-point); (2) \textbf{limiting behavior}: $\lambda \to 0$ results in uniform random, $\lambda \to \infty$ results in Nash; (3) \textbf{monotonicity}: higher $\lambda$ implies behavior that is closer to the best response; (4) \textbf{identifiability}: $\lambda$ is identified from choice data when actions have distinct expected payoffs, by the strict monotonicity of the logit function. Properties (1)--(3) are standard \citep{mckelvey1995,goeree2016}; (4) holds for the logit specification used here and is not always a property of QRE \citep{haile2008}. v1 presented these instead under a theorem heading, overstating their status here.
\end{remark}

\begin{remark}[Identifiability Verification]
For the logit QRE, $\lambda$ is point-identified precisely when games have (a) at least two actions with distinct expected payoff differences and also (b) non-degenerate type distributions, as the logit function is strictly monotone \citep{goeree2016}. (We note that \citet{haile2008} show \emph{general} QRE does not have empirical content without functional form restrictions; identifiability here rests on the logit specification.) Our games satisfy these conditions: SC has a 10-point payoff range with different best responses per value; RPD's per-round actions carry distinct expected payoffs under the given utility index (Section~\ref{sec:qre}); By construction, STST and Text-Dixit have non-degenerate payoff structures.
\end{remark}

\begin{remark}[Independence Structure]
\label{rem:iid}
The i.i.d.\ assumption in convergence theorems applies to \emph{game-level} outcomes: every complete game between a model pair produces an independent win/loss result. Within-game turn-level non-stationarity does not violate the assumption, as the Elo system consumes the final game outcome only. Wald--Wolfowitz runs tests on game-level win/loss sequences across all model pairings confirm no evidence of a serial dependence (all $p > 0.05$). We use runs test rather than Durbin-Watson because game outcomes are binary (win/loss), and DW has limited power and inapplicable asymptotics when used for discrete sequences with small samples.
\end{remark}

\section{Complete Proofs}
\label{sec:proofs}

\subsection{Proposition~\ref{thm:msne}: Exact Solve and Verification}

The Strategic Claim stage game $\Gamma_{SC}$ is a finite Bayesian game;  a mixed-strategy Bayesian Nash equilibrium exists as follows from \citet{nash1950} via \citeauthor{harsanyi1967}'s \citeyearpar{harsanyi1967} transformation.

\textbf{Computation.} The Harsanyi extensive form is encoded directly (a chance move deals $(v_1, v_2) \in \{1,\dots,6\}^2$ uniformly; each player may observe only their own value; both the claim and threshold are chosen simultaneously in a joint action $(c, t)$ with $c \in \{v,\dots,6\}$, $t \in \{1,\dots,7\}$) and solve by sequence-form linear complementarity (Gambit 16.7, \texttt{lcp\_solve}) \citep{gambit}. The solver output is cross-validated in two different ways: the Gambit expected payoff is matched to an independent analytic calculation of the behavioral-strategy payoff to $10^{-8}$, and the profile's exploitability---its best-response value minus equilibrium value, computed using an independent exact best-response routine over the factorized payoff---is zero to machine precision ($< 10^{-15}$). The solver pipeline is validated on games checkable by hand (matching pennies, rock--paper--scissors, and a dominance sanity check). Code: \texttt{scripts/arxiv\_v2\_solve\_sc.py}, to be included in the future solver and in the correction-pipeline release.

\textbf{Hand verification of the equilibrium's Bayesian consistency.} Under the equilibrium claim distribution, claim level $6$ is produced by bluffing $v=1$ types (probability $\frac{1}{6} \cdot \frac{1}{2} = \frac{1}{12}$) and honest $v=6$ types ($\frac{1}{6}$), giving the associated posterior $P(\text{bluff} \mid c\!=\!6) = \frac{1/12}{1/12 + 1/6} = \frac{1}{3}$. The same computation at claim levels $3$, $4$, and $5$ also yields precisely $\frac{1}{3}$. Under the as-implemented challenge payoffs the challenger's gain expectation from challenging a claim with bluff posterior $p$ is $4p - 2(1-p)$, which vanishes precisely at $p = \frac{1}{3}$: challengers are indifferent, supporting the mixed threshold. The equilibrium value is $11/8$ per round per player.

\textbf{Conditional versus overall bluff rates.} Types $v \in \{1, 2\}$ bluff always and $v = 3$ never does, meaning that the conditional bluff rate given $v \leq 3$ is $\beta_{\mathrm{MSNE}} = \frac{1}{3}(1 + 1 + 0) = \frac{2}{3}$, and the overall bluff rate across all values is $\frac{1}{3} \approx 0.333$. The v1 manuscript's earlier reference rate $\beta^* = 0.340$---derived from a particular stylized payoff model with challenger indifference at $P(\text{bluff} \mid c) = 2/5$---is numerically close to the equilibrium's \emph{overall} rate but is a \emph{conditional} quantity; thus they are fundamentally different objects that do not suggest convergence evidence, despite the near-coincidence (see Section~\ref{sec:results}).

\textbf{Exploitability of v1 candidate profile.} Evaluated using the implemented payoffs, the original v1 profile (bluff to $c = 6$ with probability $2/(8-v)$ for $v \leq 3$; deterministic threshold $t = 5$) yields a symmetric value of $\approx 1.397$ and admits a best response worth about $1.05$ additional points each round---it is not close to equilibrium in the game actually played by the models. \qed

\subsection{Proof of Proposition~\ref{thm:sample} (Pairwise Win-Rate Hoeffding Bound)}

\textbf{Setup.} Game outcomes $S_1, \ldots, S_n \in \{0, 1\}$ between models $i$ and $j$ are i.i.d.\ Bernoulli($p^*$), where $p^* = 1/(1 + 10^{-\Delta^*/400})$ is the real win probability under the Bradley--Terry model. The i.i.d.\ assumption at game level is validated through runs tests on game-level win/loss sequences (Remark~\ref{rem:iid}).

\textbf{Hoeffding application.} As $S_t \in [0,1]$ are i.i.d., by Hoeffding's inequality \citep{hoeffding1963}:
\begin{equation}
P(|\hat{p}_n - p^*| > \epsilon) \leq 2\exp(-2n\epsilon^2)
\end{equation}

\textbf{Conversion to Elo.} A difference in rating of $\Delta$ Elo points corresponds to a win probability $p = 1/(1 + 10^{-\Delta/400})$. Near equal strength ($\Delta = 0$, $p = 0.5$), linearization gives $\Delta \approx 694 \cdot (p - 0.5)$. A 50-point difference in Elo corresponds to $p^* \approx 0.572$, requiring detecting a deviation $\epsilon = 0.072$ from $p = 0.5$.

\textbf{Sample complexity.} Setting $2\exp(-2n \cdot 0.072^2) \leq 0.05$ gives $n \geq \lceil \ln(40)/(2 \cdot 0.005184) \rceil = 356$ games per pairing for finding 50-point differences from solely win rates. Our design makes use of 10--18 replications per pairing but achieves its ultimate model-level precision through aggregation across 28 pairings (1,855 total games, 100+ per model). Non-parametric bootstrap (Section~\ref{sec:design}) supplies the primary uncertainty quantification, with bootstrap SD of 9--80 Elo points (median of 40). \qed

\subsection{Proof of Proposition~\ref{thm:ingame} (Within-Game Convergence)}

\textbf{Setup.} Let $\beta_t$ denote the empirical rate of bluffing at round $t$ (conditional on $v \leq 3$), and $\beta^\infty$ the fixed point of the belief--response map below (a free parameter and not an equilibrium object; v2 correction). Players update their beliefs about opponent behavior through exponential smoothing.

\textbf{Belief dynamics.} Under constant-gain exponential smoothing with rate $\eta \in (0,1)$ (a Beta--Bernoulli Bayesian update would alternatively carry a declining gain $\eta_t \propto 1/t$), each player's belief $\hat{\beta}_t$ about opponent bluffing evolves as:
\begin{equation}
\hat{\beta}_{t+1} = (1-\eta)\hat{\beta}_t + \eta \cdot \mathbf{1}[\text{opponent bluffed at } t]
\end{equation}

\textbf{Best response mapping.} Given belief $\hat{\beta}$, the best response bluffing rate is modeled as affine:
\begin{equation}
\beta(\hat{\beta}) = \beta^\infty + \kappa(\hat{\beta} - \beta^\infty)
\end{equation}
where $\kappa \in (-1,1)$ is the best-response slope (players partially adjust in the direction of the map's fixed point).

\textbf{Contraction property.} The composite mapping $T: \beta_t \mapsto \E[\beta_{t+1}|\beta_t]$ satisfies:
\begin{equation}
|T(\beta) - \beta^\infty| \leq |1-\eta(1-\kappa)|\,|\beta - \beta^\infty|
\end{equation}
Setting $\rho = 1 - \eta(1-\kappa)$, with $|\rho| < 1$ for $\eta \in (0,1)$ and $\kappa \in (-1,1)$, and iterating, we have $|\E[\beta_t | \beta_1] - \beta^\infty| \leq |\rho|^{t-1}|\beta_1 - \beta^\infty|$ by the tower property of conditional expectations. This is a bound on expectation; the realized $\beta_t$ is stochastic and can depart from this trajectory. A high-probability pointwise bound would necessitate additional concentration arguments (e.g., stochastic approximation Lyapunov criteria) that lay beyond the scope of this model.

\begin{figure}[htbp]
\centering
\includegraphics[width=0.85\textwidth]{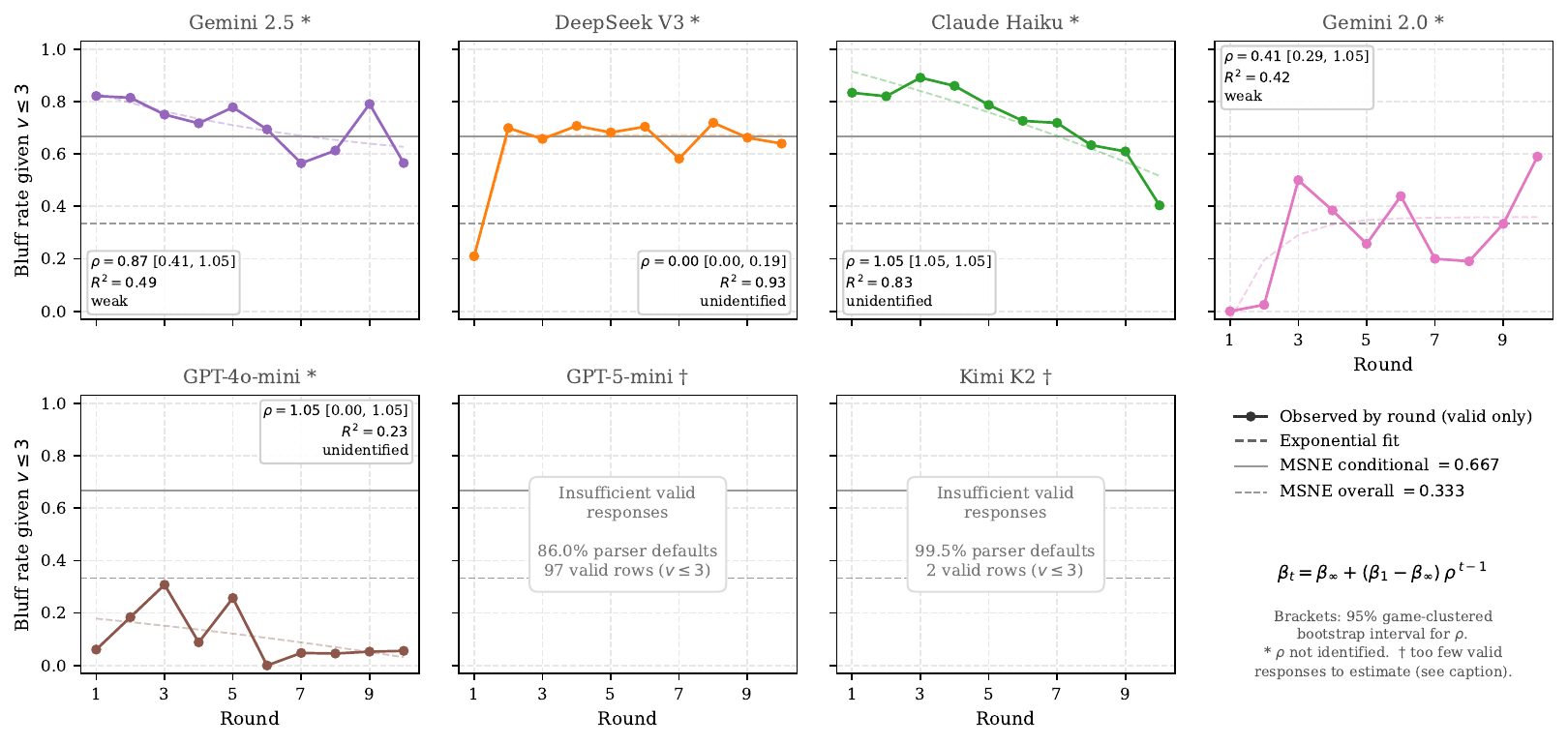}
\caption{Per-model trajectories of the conditional bluff rate (given $v \leq 3$) in Strategic Claim, on valid responses (v2). Kimi K2 and GPT-5-mini panels are excluded due to the insufficient valid data (99.5\% and 86\% parser defaults). Dashed lines demonstrate illustrative exponential-smoothing fits with a free fixed point (Proposition~\ref{thm:ingame}); game-clustered bootstrap diagnostics uncover \emph{no} model holding a well-identified convergence rate $\rho$ on valid data (fits remain weak or boundary-pinned), so no per-model $\rho$ is quoted in the text, resolving the v1 text/figure inconsistency by withdrawal. Valid-response terminal rates (rounds 9--10) range from 0.05 (GPT-4o-mini) to 0.67 (Gemini 2.5).}
\label{fig:permodel}
\end{figure}

\textbf{Empirical status (v2).} The pooled v2 block series is not monotone, so no pooled contraction factor is declared; model-specific exponential fits (with $\beta^\infty$ free) appear in Figure~\ref{fig:permodel}, and are only illustrative smoothing fits. \qed

\section{Game Specifications (as implemented; corrected in v2)}

These paragraphs describe the games as run by the engine; in places v1 stated otherwise, any discrepancy is noted. Full engine-derived fact sheets (players, action spaces, information sets, termination, scoring, payoff constants, and verbatim prompt templates) are included in the Zenodo data deposit (\texttt{reports/game\_specs.md}).

\textbf{Strategic Claim} runs for 10 rounds with a horizon known to the players (prompt shows ``Round $n$ of 10''). Each player receives a private value drawn uniformly from $\{1,...,6\}$, claims a value, and decides on a challenge threshold from 1 to 7 in the same response. Per-round payoffs result from the claim comparison ($+3$ higher / $+1$ tie / $0$ lower) plus the challenge resolution (catch a bluff: $+4$ challenger, $-4$ bluffer; false accusation: $-2$ challenger, $+2$ accused), as in Definition~\ref{def:sc}; any claimed value is never directly paid out (v1 stated otherwise). The instruction ``you cannot claim lower than your value'' is not mechanically enforced by the engine. Rows with claims below value make up 25.9\% of the log, but almost all of these are simply the parser default (claim $= 1$, threshold $= 5$) substituted for empty responses (data-validity paragraph, Methodology); real chosen under-claims by responding models are quite rare (3.2\% for DeepSeek V3, $\leq 0.5\%$ elsewhere) and are scored as honest.

\textbf{Repeated Prisoner's Dilemma} ran a fixed 15-round horizon, obscured from players (the prompt shows only the current round number; v1's ``hidden horizon drawn uniformly from 7 to 15 with commonly known upper bound'' was not implemented by the engine). Stage payoffs follow $(T,R,P,S) = (5,3,1,0)$. Players may exchange natural language messages before every decision round, which enables the game's cheap talk.

\textbf{Say the Same Thing} allows up to 20 rounds of simultaneous selection of words with a disclosed horizon. Players start with different words and must converge on an exact string match. Both players receive the identical score ($100 + 5 \cdot (20 - t)$ on a convergence at round $t$; otherwise a distance-based consolation score), so the game carries no between-model information for rating (Appendix~\ref{sec:stst_diag}). The word-distance quantity recorded as ``semantic distance'' is, as run, a character-set Jaccard fallback and not the documented embedding distance.

\textbf{Text-Dixit} is made up of 4 rounds with alternating role assignments of storyteller and guesser. Each round presents 6 AI-generated surreal scenes: one target and five decoys. The storyteller must select a clue of between 2 and 4 words from a given bank and also predicts the guesser's confidence (0--100); the guesser chooses a scene and reports confidence. Scoring: the guesser receives $+2$ for a correct identification; a calibration match (a prediction within 10 points of the guesser's reported confidence) pays the \emph{storyteller only} $+1$---although the in-game message told players ``both +1.''

\textbf{Sealed-bid first-price auction (control).} Each auction game runs 10 rounds with a disclosed horizon; private values are sampled independently each round from $\mathrm{Uniform}[0, 100]$; bids are continuous on $[0, 100]$. The winner pays their own bid, receiving $v - b$ (not floored at zero); the loser earns nothing; ties are resolved by fair coin. Players do not get information about the opponent's bids, values, or payoffs at any point, so the game is mechanically ten independent one-shot auctions---which strengthens its role as a no-opponent-modeling control (the symmetric BNE is $b(v) = v/2$). \emph{v2 note:} unparseable responses were replaced by the fallback bid of $v/2$, which is equal to the equilibrium bid, so a non-responding model scored as optimally calibrated; 88.8\% of Kimi K2's auction responses were not usable, and Gemini 2.5 played no auction games (290 games, 6 models). A secondary fallback (a bare-number regex applied when the JSON extraction fails) instead recorded the first number appearing in prose as the bid, overwriting 1--11\% of Claude Haiku's, GPT-4o-mini's, and Gemini 2.0's bids; the validity mask filters out these rows. This compromises the control's value evidentially (Section~\ref{sec:results}).

\section{Example Game Prompts}

This section provides representative prompts for two of the utilized games. Prompts are designed to elicit strategic reasoning and to remain interpretable across models.

\subsection{Text-Dixit: Storyteller Prompt}

\begin{quote}
\small
\texttt{You are playing Text-Dixit as the STORYTELLER.}

\texttt{TARGET SCENE (only you can see this):}\\
\texttt{Scene 3: "A grandfather clock melting over a park bench while chess pieces march in formation toward a door made of water."}

\texttt{ALL SCENES (your partner sees these):}\\
\texttt{1. "A library where books fly between shelves pursued by a cat wearing a monocle."}\\
\texttt{2. "An upside-down city reflected in a teacup held by enormous hands."}\\
\texttt{3. "A grandfather clock melting over a park bench while chess pieces march..."}\\
\texttt{4. "A staircase spiraling into clouds where fish swim through doorways."}\\
\texttt{5. "A violin playing itself in a garden of glass flowers under two moons."}\\
\texttt{6. "A train station where shadows travel independently of their owners."}

\texttt{WORD BANK: [time, melting, march, game, bench, water, old, movement, liquid, sitting, door, formation, clock, park, chess]}

\texttt{Your task:}\\
\texttt{1. Select 2-4 words from the word bank as your CLUE}\\
\texttt{2. Predict how confident (0-100\%) your partner will be in identifying the correct scene}

\texttt{Consider: A clue that's too obvious helps your partner but may be too easy. A clue that's too obscure may confuse them. Your score depends on accurately predicting their confidence level.}

\texttt{Respond in JSON: \{"clue": [...], "predicted\_confidence": N\}}
\end{quote}

\subsection{Text-Dixit: Guesser Prompt}

\begin{quote}
\small
\texttt{You are playing Text-Dixit as the GUESSER.}

\texttt{Your partner (the Storyteller) has a TARGET scene that only they can see. They've given you a clue to help you identify it.}

\texttt{CLUE FROM STORYTELLER: ["time", "march"]}

\texttt{SCENES:}\\
\texttt{1. "A library where books fly between shelves pursued by a cat wearing a monocle."}\\
\texttt{2. "An upside-down city reflected in a teacup held by enormous hands."}\\
\texttt{3. "A grandfather clock melting over a park bench while chess pieces march..."}\\
\texttt{4. "A staircase spiraling into clouds where fish swim through doorways."}\\
\texttt{5. "A violin playing itself in a garden of glass flowers under two moons."}\\
\texttt{6. "A train station where shadows travel independently of their owners."}

\texttt{Your task:}\\
\texttt{1. Select which scene (1-6) you believe is the target}\\
\texttt{2. Report your confidence (0-100\%) that your guess is correct}

\texttt{Respond in JSON: \{"guess": N, "confidence": N\}}
\end{quote}

\subsection{Strategic Claim: Round Prompt (verbatim engine template, v2)}

The v1 version of the appendix showed an illustrative prompt whose payoff arithmetic did not properly match the engine or Definition~\ref{def:sc}. The template below is the payoff-relevant part of the actual prompt, verbatim from the engine (\texttt{strategic\_claim.py}); verbatim templates from all five implemented games are collected and included in the Zenodo data deposit (\texttt{reports/game\_specs.md}).

\begin{quote}
\small
\texttt{Round \{n\} of \{max\_rounds\}}

\texttt{Your private value this round: \{v\}}

\texttt{[history and current scores]}

\texttt{=== GAME RULES ===}\\
\texttt{1. CLAIM: Choose a value to claim (1-6). You may claim your actual value or bluff higher.}\\
\texttt{~~~- You CANNOT claim lower than your actual value.}

\texttt{2. CHALLENGE THRESHOLD: Set a threshold for challenging opponent's claim.}\\
\texttt{~~~- You will challenge if opponent's claim is >= your threshold.}\\
\texttt{~~~- Set threshold to 7 to never challenge (no claim can be 7).}\\
\texttt{~~~- Set threshold to 1 to always challenge.}

\texttt{=== PAYOFFS ===}\\
\texttt{- Claim comparison: Higher claim wins +3, tie +1, lower +0}\\
\texttt{- Catching a bluff (opponent claimed > actual): You +4, they -4}\\
\texttt{- False accusation (opponent was truthful): You -2, they +2}

\texttt{Respond with JSON: \{"claim": <1-6>, "challenge\_threshold": <1-7>, "reasoning": "..."\}}
\end{quote}

As noted earlier, the ``cannot claim lower'' instruction is not mechanically enforced by the engine; rows with claims that fall below value (25.9\% of the log) are almost entirely defaults for empty responses from the parser rather than real chosen under-claims, and are scored as honest (Section~\ref{sec:results}).

\section{Supplementary Tables and Figures}

\begin{table}[htbp]
\centering
\small
\begin{tabular}{@{}lll@{}}
\toprule
Model & Provider & Type \\
\midrule
GPT-5-mini & OpenAI & Reasoning \\
GPT-4o-mini & OpenAI & Efficient \\
Claude Haiku 4.5 & Anthropic & Efficient \\
DeepSeek V3 & DeepSeek & Open-weight MoE \\
Kimi K2 Thinking & Moonshot & Extended reasoning \\
Gemini 2.5 Flash & Google & Efficient multimodal \\
Gemini 2.0 Flash & Google & Previous-gen efficient \\
\bottomrule
\end{tabular}
\caption{Model specifications. GPT-4o-mini was retired by OpenAI after data collection. Gemini 2.0 Flash is deprecated by Google (shutdown March 2026).}
\label{tab:models}
\end{table}

\begin{table}[htbp]
\centering
\small
\begin{tabular}{@{}lccccc@{}}
\toprule
Model & ESM & RSR & RSM & Control & SCG \\
\midrule
Claude Haiku & 100.0 & 99.9 & 100.0 & 99.0 & 100.0 \\
DeepSeek V3 & 100.0 & 100.0 & 100.0 & 100.0 & 99.9 \\
GPT-4o-mini & 100.0 & 100.0 & 100.0 & 89.9 & 99.8 \\
GPT-5-mini & 76.4 & 14.0$^{d}$ & 99.0 & 100.0 & 97.3 \\
Gemini 2.0 & 100.0 & 100.0 & 100.0 & 88.9 & 100.0 \\
Gemini 2.5 & 35.3$^{d}$ & 80.5 & 77.4 & --- & 18.0$^{d}$ \\
Kimi K2 & 14.0$^{d}$ & 0.5$^{d}$ & 21.7$^{d}$ & 11.4$^{d}$ & 17.0$^{d}$ \\
\bottomrule
\end{tabular}
\caption{Percent of acting moves that remained valid under the per-move validity mask (moves where recorded action matches the action recoverable from the raw response), by model and game. $^{d}$ marks cells that fall below the 50\% retention threshold, dropped from rating analyses (Table~\ref{tab:elo}); Gemini 2.5 did not play auction games. Full disjoint counts (empty, truncated-unrecoverable, overwritten, and recovered) are compiled and included in the Zenodo data deposit (\texttt{reports/validity\_share\_table.md}).}
\label{tab:validity}
\end{table}

\begin{table}[htbp]
\centering
\small
\begin{tabular}{@{}lcccc@{}}
\toprule
& \multicolumn{2}{c}{Valid responses} & \multicolumn{2}{c}{As recorded (incl.\ defaults)} \\
\cmidrule(lr){2-3} \cmidrule(lr){4-5}
Round & $\beta$ ($n$) & Coop.\ ($n$) & $\beta$ ($n$) & Coop.\ ($n$) \\
\midrule
1-2 & 0.565 $\pm$ 0.020 (623) & 0.945 $\pm$ 0.006 (1518) & 0.429 (820) & 0.897 (1600) \\
3-4 & 0.638 $\pm$ 0.020 (555) & 0.954 $\pm$ 0.006 (1441) & 0.451 (785) & 0.859 (1600) \\
5-6 & 0.602 $\pm$ 0.021 (545) & 0.938 $\pm$ 0.006 (1398) & 0.414 (792) & 0.820 (1600) \\
7-8 & 0.483 $\pm$ 0.022 (534) & 0.930 $\pm$ 0.007 (1368) & 0.330 (781) & 0.795 (1600) \\
9-10 & 0.515 $\pm$ 0.022 (538) & 0.894 $\pm$ 0.008 (1361) & 0.344 (806) & 0.761 (1600) \\
11-15 & --- & 0.895 $\pm$ 0.005 (3397) & --- & 0.760 (4000) \\
\bottomrule
\end{tabular}
\caption{Behavioral dynamics per round block (v2; $\pm$ binomial SE). SC bluffing rate is conditional on $v \leq 3$. \emph{Valid responses} restricts to only those rows whose raw model response is not empty; \emph{as recorded} includes the parser defaults that were substituted for empty responses (claim $= 1$/never bluff in SC; defect in RPD), which depress both series. References (Proposition~\ref{thm:msne}): MSNE conditional rate $2/3$; MSNE overall rate $1/3$; uniform baseline $0.5$. No equilibrium benchmark attached to cooperating (hidden fixed horizon; Proposition~\ref{thm:folk}). The v1 version of this table ($0.52 \to 0.35$) did not reproduce under either construction (change log).}
\label{tab:convergence}
\end{table}

\begin{table}[htbp]
\centering
\small
\setlength{\tabcolsep}{4pt}
\begin{tabular}{@{}llp{0.47\textwidth}@{}}
\toprule
Quantity & Source & Result (v2) \\
\midrule
MSNE conditional bluff rate $2/3$ & Prop.~\ref{thm:msne} & valid-response rounds 9--10: $0.515 \pm 0.022$ \\
MSNE overall bluff rate $1/3$ & Prop.~\ref{thm:msne} & near the \emph{as-recorded} 0.344, which parser defaults dilute \\
Cooperation (behavioral) & Prop.~\ref{thm:folk} & block means 0.89--0.95 among valid responses (no equil.\ benchmark) \\
Smoothing-model fit & Prop.~\ref{thm:ingame} & illustrative only; pooled series non-monotone \\
Aggregated rating precision & Prop.~\ref{thm:sample}; bootstrap & bootstrap SD 9--80 (median 40) \\
Axis structure & Design & unanswerable after validity masking ($n = 4$--$5$ valid cells; Section~\ref{sec:results}) \\
\bottomrule
\end{tabular}
\caption{Framework quantities and corrected empirical results (v2). The original v1 presentation of this table scored ``$\beta^* \to 0.340$: within 4\%'' as a confirmed prediction; the exact solve shows that comparison conflated the equilibrium's overall and conditional rates of bluffing (Proposition~\ref{thm:msne}).}
\label{tab:predictions}
\end{table}

\begin{table}[htbp]
\centering
\small
\begin{tabular}{@{}lrrcr@{}}
\toprule
Model & $n$ & $\hat{\lambda}$ & 95\% CI / bound & $\tilde{\lambda}$ \\
\midrule
\multicolumn{5}{@{}l}{\textit{Strategic Claim}} \\
GPT-4o-mini & 700 & 1.11 & [0.93, 1.31] & 3.80 \\
Gemini 2.0 & 700 & 0.85 & [0.73, 1.00] & 2.94 \\
Gemini 2.5$^{r}$ & 1100 & 0.77 & [0.70, 0.86] & 2.63 \\
DeepSeek V3 & 1100 & 0.62 & [0.56, 0.68] & 2.07 \\
Claude Haiku & 1100 & 0.53 & [0.47, 0.58] & 1.77 \\
Kimi K2 & --- & \multicolumn{3}{l}{not estimable: 99.5\% parser defaults} \\
GPT-5-mini & --- & \multicolumn{3}{l}{not estimable: 86\% parser defaults} \\
\addlinespace
\multicolumn{5}{@{}l}{\textit{Repeated Prisoner's Dilemma}} \\
GPT-4o-mini & 1050 & 0.00 & $\le$ 0.002 & 0.00 \\
Gemini 2.0 & 1050 & 0.00 & $\le$ 0.002 & 0.00 \\
Gemini 2.5$^{r}$ & 1650 & 0.00 & $\le$ 0.002 & 0.00 \\
DeepSeek V3 & 1650 & 0.00 & $\le$ 0.001 & 0.00 \\
Claude Haiku & 1650 & 0.00 & $\le$ 0.001 & 0.00 \\
GPT-5-mini & 1650 & 0.00 & $\le$ 0.001 & 0.00 \\
Kimi K2 & --- & \multicolumn{3}{l}{retracted: 78\% forced-defect defaults (v1: 1.10)} \\
\bottomrule
\end{tabular}
\caption{QRE $\hat{\lambda}$ estimates, re-calculated for v2 on the entire legal claim set $\{1,\dots,6\}$ with complete rows, restricted to models with real responses (v2 data-validity audit; change log). Cells for Kimi K2 (SC, RPD) and GPT-5-mini (SC) are not estimable: their rows are dominated by the parser defaults switched for empty responses (claim $= 1$/threshold $= 5$ in SC; forced defection in RPD)---the v1 values for those cells, including Kimi K2's RPD $\hat{\lambda} = 1.10$, characterized only the parser instead of the models as intended (valid-response shares: 0.5\% and 14\% in SC; 22\% in Kimi's RPD). $^{r}$Gemini 2.5's responses were truncated and its actions regex-recovered from reasoning text; treat with reservations. $\hat{\lambda}$ is the grid MLE on $[0,5]$; intervals are full-game cluster-bootstrap 95\% percentiles ($B = 2000$); boundary cells record one-sided 95\% profile-likelihood upper bounds (Chernoff mixture). $\tilde{\lambda} = \hat{\lambda} \cdot \overline{\Delta U}$ is the payoff-normalized index (Section~\ref{sec:qre}). A human-range comparison is not made.}
\label{tab:qre_cv}
\end{table}

\begin{table}[htbp]
\centering
\small
\begin{tabular}{@{}lccccc@{}}
\toprule
Pair & $n$ & $r$ (BT) & Bootstrap 95\% CI & Holm $p$ (exact perm.) & v1 (retracted inference) \\
\midrule
\multicolumn{6}{@{}l}{\textit{As recorded (includes validity-contaminated cells)}} \\
ESM--RSR & 7 & $-0.81$ & $[-0.93, -0.54]$ & 0.046 & $-0.95$, ``$p=0.0009$'' \\
RSR--RSM & 7 & $-0.82$ & $[-0.89, -0.70]$ & 0.018 & $-0.82$, $p=0.02$ \\
ESM--RSM & 7 & $+0.39$ & $[+0.07, +0.62]$ & 0.35 & $+0.72$, $p=0.07$ \\
\addlinespace
\multicolumn{6}{@{}l}{\textit{Validity-masked (valid model-axis cells only; permutation floor $2/n!$)}} \\
ESM--RSR & 4 & $-0.36$ & $[-0.96, +0.85]$ & 1.0 (floor 0.083) & --- \\
RSR--RSM & 5 & $+0.04$ & $[-0.70, +0.76]$ & 1.0 (floor 0.017) & --- \\
ESM--RSM & 5 & $+0.89$ & $[+0.74, +0.96]$ & 0.25 (floor 0.017) & --- \\
\addlinespace
\multicolumn{6}{@{}l}{\textit{Auction control (as recorded; evidential value compromised)}} \\
Control--ESM & 6 & $+0.36$ & $[+0.02, +0.63]$ & n.s. & ``$+0.09$'' (non-reproducible) \\
Control--RSR & 6 & $-0.41$ & $[-0.61, -0.19]$ & n.s. & ``$+0.11$'' (non-reproducible) \\
Control--RSM & 6 & $+0.34$ & $[+0.17, +0.50]$ & n.s. & ``$+0.14$'' (non-reproducible) \\
\bottomrule
\end{tabular}
\caption{Inter-axis correlations with fixed inference (v2): batch Bradley--Terry ratings, full-refit game-level bootstrap ($B = 5000$), Holm-corrected exact permutation tests. The as-recorded block amends the v1 estimator and inference but still includes parser-influenced cells; the validity-masked block is definitive and shows that cross-axis questions are \emph{unanswerable} at the remaining $n$ (at $n = 4$ the smallest attainable two-sided permutation $p$ is 0.083). The bluff-disposition partial correlation on as-recorded data is $r(\text{ESM},\text{RSR} \mid \beta) = -0.80$ (CI excluding zero); on valid data it has one residual degree of freedom and retains no inferential weight. Control rows: $n = 6$ (Gemini 2.5 played no auctions); the v1 control row did not reproduce under various constructions, and 88.8\% of Kimi K2's auction bids were simply parser-imputed equilibrium bids. SCG omitted (Appendix~\ref{sec:stst_diag}).}
\label{tab:corr}
\end{table}

\subsection{Temperature Ablation}

To assess whether QRE $\lambda$ estimates are confounded by LLM sampling temperature, we run 10 self-play Strategic Claim games on each temperature $T \in \{0.3, 0.5, 0.7, 1.0\}$ with Claude Haiku (40 games total).

\begin{table}[htbp]
\centering
\small
\begin{tabular}{@{}ccc@{}}
\toprule
$T$ & $N$ & Bluff rate $\beta$ ($\pm$SE) \\
\midrule
0.3 & 10 & $0.570 \pm 0.050$ \\
0.5 & 10 & $0.645 \pm 0.046$ \\
0.7 & 10 & $0.495 \pm 0.049$ \\
1.0 & 10 & $0.554 \pm 0.049$ \\
\bottomrule
\end{tabular}
\caption{Temperature ablation results from Claude Haiku self-play in Strategic Claim (10 games at each temperature, 10 rounds each). Bluff rate $\beta$ is conditional on $v \leq 3$. With 10 games in each cell the ablation is underpowered; the non-monotone variation with $T$ is consistent with noise and cannot resolve the $\lambda$/temperature confound, which would require a token-level logit access.}
\label{tab:temp_ablation}
\end{table}

\subsection{Prior Sensitivity Analysis}

To verify that Bayesian $\lambda$ estimates are indeed robust to prior specification, we compare the posterior means under three different Gamma priors: Gamma(1, 0.5) (diffuse), Gamma(2, 1) (weakly informative; default), and Gamma(3, 1) (moderately informative). Table~\ref{tab:prior_sens} reports the posterior means and the range across priors for every model.

\begin{table}[htbp]
\centering
\small
\begin{tabular}{@{}lcccc@{}}
\toprule
Model & Gamma(1, 0.5) & Gamma(2, 1) & Gamma(3, 1) & Range \\
\midrule
GPT-4o-mini & 0.603 & 0.606 & 0.609 & 0.006 \\
Gemini 2.0 & 0.382 & 0.385 & 0.390 & 0.008 \\
Gemini 2.5 & 0.323 & 0.326 & 0.329 & 0.006 \\
DeepSeek V3 & 0.216 & 0.220 & 0.224 & 0.008 \\
Kimi K2 & 0.033 & 0.059 & 0.082 & 0.050 \\
Claude Haiku & 0.032 & 0.047 & 0.058 & 0.026 \\
GPT-5-mini & 0.030 & 0.054 & 0.074 & 0.043 \\
\bottomrule
\end{tabular}
\caption{Prior sensitivity analysis for the original v1 Bayesian QRE layer (Strategic Claim), kept for the record. Maximum posterior-mean variation across the three Gamma priors is $< 0.01$ for models with moderate $\lambda$ and $\leq 0.05$ for near-boundary estimates. The Kimi K2 and GPT-5-mini rows are contaminated by the parser behavior and carry no discernible behavioral meaning (their SC cells are not estimable in v2; Table~\ref{tab:qre_cv}). Note (v2): these posteriors were computed on the v1 action set, which dropped all under-claim rounds (change log); the v2 uncertainty quantification is via the cluster bootstrap and one-sided bounds of Table~\ref{tab:qre_cv}.}
\label{tab:prior_sens}
\end{table}

\subsection{Model Expansion Studies}

To assess framework generalizability beyond the original seven evaluated models and temporal stability of QRE rankings, we conduct two expansion studies. \textbf{B1 (New Families)} examines two models from providers not in the original ensemble: Grok 4.1 Fast (xAI) and MiniMax M2.5 (MiniMax), each playing 20 SC self-play games. \textbf{B2 (Version Stability)} tests the successors of two originally evaluated models, DeepSeek V3.2 (successor to V3) and Kimi K2.5 (successor to K2 Thinking), under identical conditions (20 games each; 80 games across both studies).

\begin{table}[htbp]
\centering
\small
\begin{tabular}{@{}llcccl@{}}
\toprule
Study & Model & $N$ & $\beta$ ($\pm$SE) & $\hat{\lambda}$ & Notes \\
\midrule
B1 & Grok 4.1 Fast & 20 & $0.473 \pm 0.037$ & 0.30 & New family \\
B1 & MiniMax M2.5 & 20 & $0.498 \pm 0.034$ & 0.00 & New family \\
\midrule
B2 & DeepSeek V3.2 & 20 & $0.288 \pm 0.032$ & 0.70 & cf.\ V3: $\beta\!=\!0.523$, $\lambda\!=\!0.2$ \\
B2 & Kimi K2.5 & 20 & $0.292 \pm 0.033$ & 0.60 & cf.\ K2: $\beta\!=\!0.173$, $\lambda\!=\!0.8$ \\
\bottomrule
\end{tabular}
\caption{Model expansion results (SC self-play, 20 games each). The ``cf.''\ baselines in the Notes column are v1 self-play-pipeline values; they are not comparable to Table~\ref{tab:qre_cv}. Response-validity addendum (v2): the expansion runs are clean (0--2.5\% empty responses; Kimi K2.5: 0\%), so these values reflect real behavior---notably, K2.5 \emph{answers} where the original K2 was unable to within the token budget. For that same reason the B2 ``successor vs.\ original'' deltas involving K2 cannot be interpreted as capability trajectories: the original-K2 baseline is highly contaminated by the parser (Data validity, Methodology). DeepSeek V3.2 vs.\ V3 stands with the usual self-play caveat. Expansion studies employ self-play due to cost efficiency; cross-model pairings may produce other dynamics.}
\label{tab:model_expansion}
\end{table}

\subsection{Behavioral-Model Comparison on a Common Event Space}
\label{sec:modelcomp}

The original v1 version of this subsection reported a comparison of BIC (``Table 12'') where the likelihoods were evaluated on event spaces that were model-dependent and over parse-dependent row sets: the implied uniform action-space size, which can be recovered from the published numbers as $\exp(\mathrm{BIC}_{\mathrm{Rand}}/2n)$, ranged from 2.9 to 5.9 between models, and the per-model row counts were between 175 and 1{,}099 where complete records contain 800--1{,}280. Comparisons across models on that table were not valid, and the three-cluster claim built using it (near-Nash / QRE-rational / near-random) is retracted. The comparison below succeeds it: all candidates are scored on a singular event space (the binary bluff/honest decision at $v \leq 3$) over complete rows constructed identically, with a symmetric decision rule and game-level held-out evaluation.

\begin{table}[htbp]
\centering
\footnotesize
\setlength{\tabcolsep}{3pt}
\begin{tabular}{@{}lrrrrrrrrl@{}}
\toprule
 & & & \multicolumn{6}{c}{Held-out mean log-likelihood per observation} & \\
\cmidrule(lr){4-9}
Model & $n$ & $\hat{\beta}$ & Uniform & Mixture & $\beta^\star$-ref & MSNE-ref & MSNE-marg & QRE & Verdict \\
\midrule
Claude Haiku & 649 & 0.732 & -0.693 & -0.620 & -0.858 & -7.191 & -0.615 & -0.693 & Mixture $\approx$ MSNE-marg \\
DeepSeek V3 & 648 & 0.630 & -0.693 & -0.650 & -0.857 & -6.410 & -0.644 & -0.691 & Mixture $\approx$ MSNE-marg \\
GPT-4o-mini & 375 & 0.109 & -0.693 & \textbf{-0.363} & -0.472 & -10.032 & -1.020 & -0.693 & Mixture ($\geq 3.2$\,SE) \\
Gemini 2.0 & 395 & 0.291 & -0.693 & \textbf{-0.605} & -0.619 & -7.645 & -0.899 & -0.693 & Mixture ($\geq 4.0$\,SE) \\
Gemini 2.5$^{r}$ & 629 & 0.711 & -0.693 & \textbf{-0.608} & -0.887 & -6.156 & -0.611 & -0.680 & Mixture ($\geq 2.1$\,SE) \\
GPT-5-mini$^{\dagger}$ & 638 & 0.127 & -0.693 & \textbf{-0.365} & -0.497 & -8.631 & -1.016 & -0.693 & (parser-dominated) \\
Kimi K2$^{\dagger}$ & 650 & 0.003 & -0.693 & \textbf{-0.039} & -0.423 & -8.943 & -1.096 & -0.693 & (parser-dominated) \\
\bottomrule
\end{tabular}
\caption{Behavioral-model comparison on common event space, replacing the now retracted v1 BIC table. The event is the bluff/honest decision on each Strategic Claim round with $v \leq 3$, one row per player-round, from the engine's outcome records, so all seven models are scored on rows constructed identically ($n$ differs only in game counts, not in parse success). Entries are held-out mean log-likelihood per observation across five game-level 70/30 splits (seeds 1--5) stratified by pairing; rounds in every game are not separated. Candidates: Uniform ($P(\text{bluff}) = 0.5$, $k{=}0$); Mixture (single fitted rate $p_m$, $k{=}1$); $\beta^\star$-ref ($2/(8-v)$, $k{=}0$; the v1 derived rate, a reference curve); MSNE-ref (the exact equilibrium of Proposition~\ref{thm:msne}, $k{=}0$; its degenerate predictions are floored at $10^{-6}$, so the magnitude is only readable as ``decisively rejected''); MSNE-marg (the equilibrium's \emph{aggregate} conditional rate, $P(\text{bluff} \mid v \leq 3) = 2/3$, $k{=}0$; added in v2); QRE (logistic in $\lambda(U_{\text{bluff}} - U_{\text{honest}})$, $k{=}1$). Bolding required besting every rival by 2 SE or more of the paired split-to-split difference; no candidate is able to separate for Claude Haiku or DeepSeek V3, where Mixture and MSNE-marg are indistinguishable statistically (Gemini 2.5's Mixture margin over MSNE-marg is $2.1$ SE). $^{\dagger}$Rows computed on the parser-contaminated data (99.5\% / 86\% empty-response defaults; data-validity audit); these support no behavioral claim. $^{r}$Gemini 2.5's actions are regex-recovered from truncated responses.}
\label{tab:bic}
\end{table}

\begin{figure}[htbp]
\centering
\includegraphics[width=0.85\textwidth]{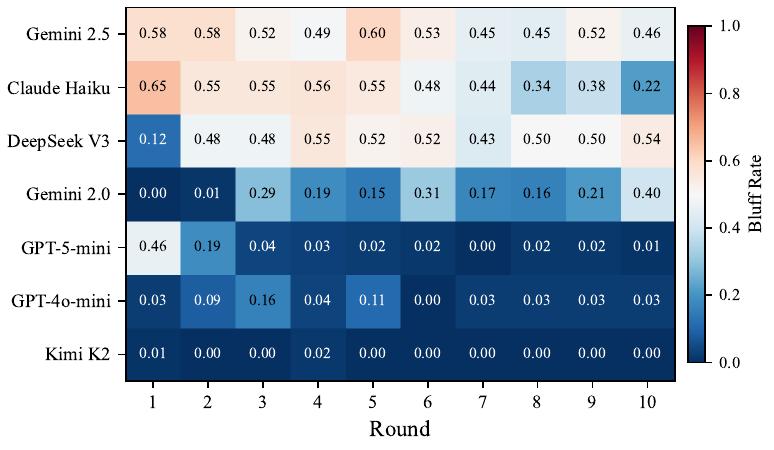}
\caption{Strategic Claim bluff rates by model and by round. Bluff likelihood varies widely across the different models, from frequent bluffers among the valid responders (Gemini 2.5, Claude Haiku, DeepSeek V3; $\beta > 0.40$) to rare bluffers (GPT-4o-mini; $\beta < 0.15$); GPT-5-mini and Kimi K2 are omitted from this comparison (parser-contaminated SC rows). These are descriptive bluff-likelihood distinctions; behavioral-model attribution is left in the common-event-space comparison (Table~\ref{tab:bic}), and low bluff rates fall outside of equilibrium play (the MSNE conditional rate is $2/3$; Proposition~\ref{thm:msne}).}
\label{fig:sc_heatmap}
\end{figure}

\paragraph{Corrected heterogeneity finding.} On held-out data, a single fitted bluff rate---a model without real strategic content---predicts behavior at least as well as every game-theoretic candidate for all models with reliable response data, and strictly better than uniformly, the v1 $\beta^\star$ reference, the equilibrium's type-level prescriptions, and the logit QRE (with a fitted $\lambda$ that collapses to zero on this event space for almost all of the models). One aggregate is able to keep pace: a constant at the equilibrium's conditional rate that is parameter-free ($2/3$; MSNE-marg, added in v2) cannot be statistically distinguished from the fitted propensity for Claude Haiku and DeepSeek V3, and within $2.1$ SE of it for Gemini 2.5---exactly the models with rates ($0.63$--$0.73$) that lie near $2/3$. The equilibrium's \emph{type-level} structure is decisively dismissed wherever testable: per-decision consensus with its deterministic prescriptions lands at around the coin-flip level (0.29--0.56). The equilibrium therefore gets an overall propensity roughly correct for some models but the conditioning on private value incorrect for every single one. What does survive is strong cross-model \emph{propensity} heterogeneity among the responding models (bluff rates 0.109--0.732; Figure~\ref{fig:sc_heatmap}) with value-conditional patterns no candidate captures (Claude Haiku instead bluffs \emph{more} at higher private values---0.50, 0.82, 0.88 for $v = 1, 2, 3$---the opposite of any equilibrium-motivated prediction). The Kimi K2 and GPT-5-mini rows are not included in any behavioral interpretation: the parser constants merely appear as ultra-low propensities, and the mixture's enormous ``win'' margins there quantify that artifact, not behavior in the model. The result is negative but usable: these games are able to elicit large, reliable behavioral differences across different responding LLMs, but attributing them to bounded-rational strategic reasoning via QRE fits is not supported by the data produced here---and for benchmark pipelines in general, response-validity auditing has to precede behavioral model comparison, as a silent parser default is indistinguishable from a very extreme behavioral type in the record of the outcomes. This sharpens the mixed-fit pattern that is reported by \citet{jia2025}: with no fitted-propensity null and without a validity mask, heterogeneous ``best fits'' among equilibrium-anchored models may only actually represent propensity differences or harness artifacts rather than serious strategic regimes, consistent with human-subject evidence that non-equilibrium behavioral models can out-predict equilibrium ones \citep{wright2017}.

\subsection{Identifiability: Profile Likelihoods and Parameter Recovery (new to v2)}
\label{sec:identifiability}

\begin{figure}[htbp]
\centering
\includegraphics[width=0.95\textwidth,height=0.82\textheight,keepaspectratio]{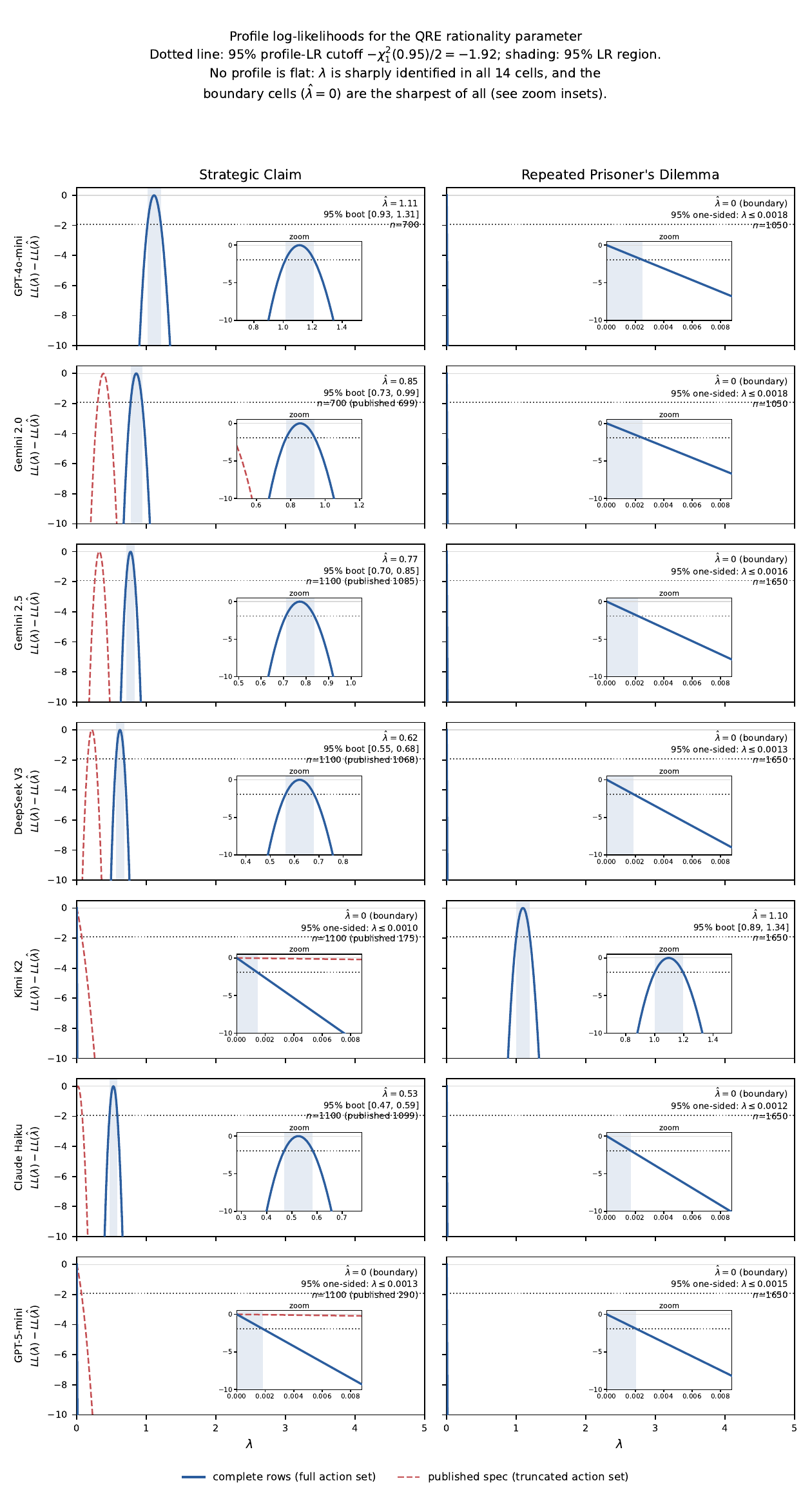}
\caption{Normalized profile log-likelihoods $\mathrm{LL}(\lambda) - \mathrm{LL}(\hat{\lambda})$ per model $\times$ game cell on complete rows, with the 95\% profile-LR cutoff present at $-1.92$. Sharply peaked curves (e.g.\ SC for GPT-4o-mini) indicate a well-identified $\hat{\lambda}$; curves that decline monotonically from $\lambda = 0$ (boundary cells) show what non-identification looks like and also motivate the one-sided bounds used in Table~\ref{tab:qre_cv}.}
\label{fig:profilell}
\end{figure}

\begin{figure}[htbp]
\centering
\includegraphics[width=0.95\textwidth]{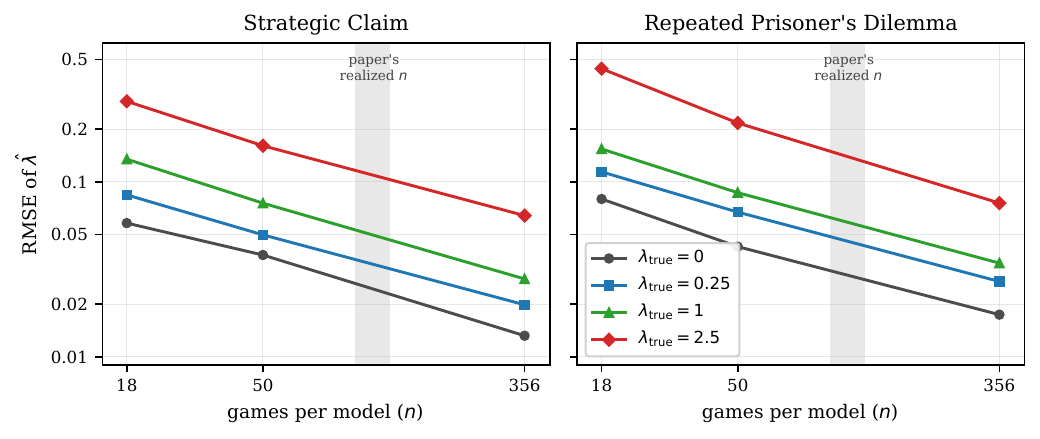}
\caption{The parameter recovery under the estimator's data-generating process (well-specified case; 200 replicates per cell, seed fixed): RMSE of $\hat{\lambda}$ versus the number of games, by $\lambda_{\text{true}} \in \{0, 0.25, 1, 2.5\}$, for SC and RPD. The shaded band marks this paper's realized $\approx$ 100--130 games per model. Recovery is in essence unbiased from $n = 18$ games on; what limits the estimator is the $\lambda = 0$ boundary (52--79\% of replicates land precisely on it, where Wald intervals are undefined), and logit saturation makes large $\lambda$ intrinsically much noisier (Fisher information per game at $\lambda = 2.5$ is $\approx$ 19\% of its $\lambda = 1$ value in SC).}
\label{fig:recovery}
\end{figure}

Together these delineate what the $\lambda$ layer is and cannot support: at the implemented sample sizes, an interior $\lambda$ is estimated accurately and with valid Wald coverage (87--100\% across recovery cells), so \emph{sample size does not act as the binding constraint}; the binding constraints are the boundary (a number of the cells sit at $\hat{\lambda} \approx 0$, where only one-sided statements can be made) as well as the validity of the utility index (Section~\ref{sec:modelcomp}). Full recovery tables: \texttt{reports/recovery.md} in the Zenodo data deposit.

\subsection{STST Diagnosis (new in v2)}
\label{sec:stst_diag}

The v1 manuscript had originally reported STST's zero variance in rating as ``focal-point coordination is trivially solved.'' The updated v2 forensic diagnosis (script \texttt{arxiv\_v2\_stst\_diag.py}) relates three facts. (1) \emph{Zero variance is essentially structural, not behavioral}: because the engine assigns both seats an identical score in every game, every game is a rating draw and all ratings are pinned at exactly 1500 for any models and any play; the $1500 \pm 0$ is simply an algebraic identity. (2) \emph{A third of the logged moves are not actually model choices}: with a 1{,}024-token completion budget, Kimi K2 and Gemini 2.5 exhausted tokens before contributing an answer in most rounds (83.5\% and 86.3\% of their STST moves unusable, respectively), and the parser recorded artifacts---the token ``word'' is the second most common ``converging word'' in the corpus (62 games). (3) \emph{The copy-strategy objection is not supported evidentially}: copying the opponent's previous word makes up only 4.4\% of moves, and mutual copying cannot produce convergence (it swaps words). On the clean subset (264 games with every move parsed), the central factor is that the task is at ceiling: 99.6\% converge, 63\% on round one, through the desired independent selection of the same focal word. The SCG axis is thus retired from rating analyses, and a potential STST redesign (asymmetric scoring, strict parsing, adequate token budgets) is future work.

\begin{figure}[htbp]
\centering
\includegraphics[width=0.85\textwidth]{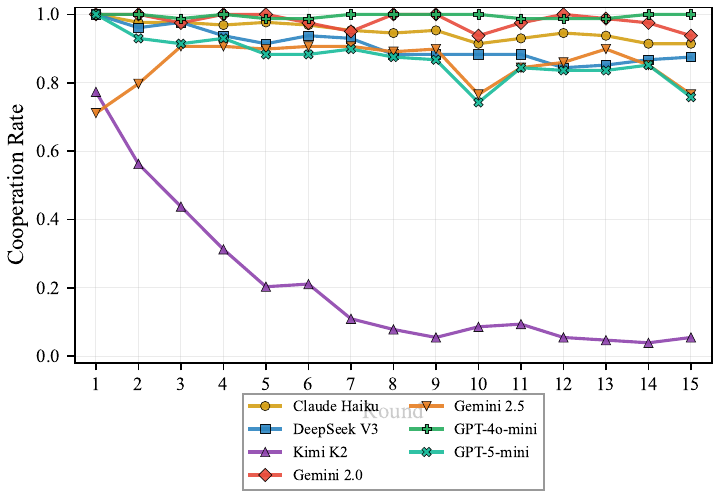}
\caption{Per-model cooperation trajectories in the Repeated Prisoner's Dilemma (15 rounds), as originally recorded. Kimi K2's apparent decline from 77\% to 5\% is merely a parser artifact---78\% of its responses were empty (with the token budget exhausted by concealed reasoning) and the parser records a defection action for any instance of an empty response; the v1 reading of this trajectory, and the RPD $\hat{\lambda} = 1.10$ built on it, are therefore withdrawn. Models with real responses sustain 85--99.5\% cooperation throughout (GPT-4o-mini: 99.5\%).}
\label{fig:rpd_cooperation}
\end{figure}

% \newpage
% \input{checklist.tex}   % NeurIPS checklist dropped for the public arXiv version (G0 Branch A)

\end{document}